\documentclass[eqsecnum,showpacs,amsmath,aps,nofootinbib]{revtex4}  
\usepackage{epsfig}
\usepackage{amssymb}
\usepackage[dvips]{color}
\usepackage[latin1]{inputenc}
\usepackage{graphicx}
\usepackage{gensymb}
\usepackage{amsmath}
\usepackage{subfig}
\usepackage{relsize}
\usepackage{mathtools}

\def\beq{\begin{equation}}
\def\eeq{\end{equation}}
\def\beqq{\begin{eqnarray}}
\def\eeqq{\end{eqnarray}}

\def\O{{\rm O}}

\newcommand{\bdm}{\begin{displaymath}}
\newcommand{\edm}{\end{displaymath}}
\newcommand{\E}{{\rm e}}

\def\pmb#1{\setbox0=\hbox{$#1$}%
  \kern-.025em\copy0\kern-\wd0
  \kern.05em\copy0\kern-\wd0
  \kern-.025em\raise.0433em\box0}

\makeatletter
\renewcommand*{\@fnsymbol}[1]{\ensuremath{\ifcase#1\or *\or \dagger\or
    \ddagger\or 
   \mathsection\or **\or \dagger\dagger
   \or \ddagger\ddagger \else\@ctrerr\fi}}
\makeatother

\begin{document}
\title{On the foundations of general relativistic celestial mechanics} 

\author{Emmanuele Battista}
\email[E-mail: ]{ebattista@na.infn.it}
\affiliation{Istituto Nazionale di Fisica Nucleare, Sezione di Napoli, Complesso Universitario di Monte
S. Angelo, Via Cintia Edificio 6, 80126 Napoli, Italy}

\author{Giampiero Esposito}
\email[E-mail: ]{gesposit@na.infn.it}
\affiliation{Istituto Nazionale di Fisica Nucleare, Sezione di
Napoli, Complesso Universitario di Monte S. Angelo, 
Via Cintia Edificio 6, 80126 Napoli, Italy}

\author{Simone Dell'Agnello}
\email[E-mail: ]{simone.dellagnello@lnf.infn.it}
\affiliation{Istituto Nazionale di Fisica Nucleare, Laboratori Nazionali di Frascati,
00044 Frascati, Italy}

\date{\today}

\begin{abstract}
Towards the end of nineteenth century, Celestial Mechanics provided the most powerful tools 
to test Newtonian gravity in the solar systems, and led also to the discovery of chaos in 
modern science. Nowadays, in light of general relativity, Celestial Mechanics leads to a
new perspective on the motion of satellites and planets. The reader is here introduced
to the modern formulation of the problem of motion, following what the leaders in the field
have been teaching since the nineties. In particular, the use of a global chart for the overall
dynamics of $N$ bodies and $N$ local charts describing the internal dynamics of each body.
The next logical step studies in detail how to split the $N$-body problem into two sub-problems 
concerning the internal and external dynamics, how to achieve the effacement properties that 
would allow a decoupling of the two sub-problems, how to define external-potential-effacing 
coordinates and how to generalize the Newtonian multipole and tidal moments. 
The review paper ends with an assessment of the nonlocal equations of motion obtained within
such a framework, a description of the modifications induced by general relativity
on the theoretical analysis of the Newtonian three-body problem, and a mention of the
potentialities of the analysis of solar-system metric data carried out with the Planetary 
Ephemeris Program.
\end{abstract}

\pacs{04.20.Cv, 95.10.Ce}

\maketitle

\section{Introduction}

At the end of the nineteenth century, two hundred years after the publication of Newton's Principia,
Celestial Mechanics was a very successful branch of Mechanics (with hindsight, we would speak of 
Classical Mechanics, but the Planck constant had not yet been postulated nor measured). The monumental
treatise by Tisserand \cite{T1,T2,T3,T4} had studied thoroughly planetary motions and their 
perturbations \cite{T1}, rotational motion of celestial bodies \cite{T2}, 
theories of lunar motion \cite{T3}, the theories of Jupiter and Saturn satellites \cite{T4}, and
perturbations of small planets \cite{T4}. Moreover, in his outstanding work on new methods in
celestial mechanics \cite{P1,P2,P3,P4}, Poincar\'e discovered periodic solutions of the three-body
problem \cite{P2,P4}, asymptotic \cite{P2} and doubly asymptotic solutions \cite{P4}, the
nonexistence of uniform integrals \cite{P2}, the theory of integral invariants \cite{P4},
and the whole topic of chaos became part of modern science, jointly with an assessment of 
asymptotic methods for studying secular terms \cite{P3} in the equations of celestial mechanics, 
gaining a better understanding of the limitations of several methods used by astronomers and applied
mathematicians.

However, as was stressed by Poincar\'e in the introduction \cite{P2} to volume 1 of his monumental
treatise, the relevance of celestial mechanics from the point of view of fundamental physics
lies mainly in the possibility of using it to ascertain whether Newton's theory provides the best
theory of gravity at all scales (i.e. within the solar system and far beyond that). Although 
mankind landed safely on the moon thanks to Newtonian celestial mechanics in Szebehely's book
\cite{Sze}, the discovery of general relativity led to a novel perspective on the classical
problems of Newtonian celestial mechanics, as is clear from the work of Lorenz and Droste \cite{Lor},
Einstein et al. \cite{E1,E2}, Levi-Civita \cite{Levi-Civita}, Fock \cite{Fock1959}, Brumberg \cite{BR1}, Kopeikin
\cite{K1}, Yamada et al. \cite{Y1}. But the most systematic application of general relativity 
methods to celestial mechanics is due to Damour and his collaborators \cite{D1987,BD,D1,D2,D3,D4} 
(cf. Ref. \cite{ALBA1}).

In our review paper, aimed at helping the general reader to become familiar with the modern formulation
of the problem of motion, Sec. II is devoted to the theory of reference systems, 
Sec. III studies Newtonian tidal and multipole expansions, Sec. IV leads to relativistic tidal and multipole 
expansions, and the monopole model is briefly reviewed in Sec. V. 
Last, the integro-differential dynamical equations are discussed in Sec. VI and the effects
of general relativity on the restricted three-body problem are summarized in Sec. VII;
Sec. VIII outlines the potentialities of the analysis of solar-system metric data carried out
with the Planetary Ephemeris Program, while concluding remarks are presented in Sec. IX.

\section{Theory of reference systems}

We here outline the formalism lying behind general relativistic celestial 
mechanics of systems of $N$ arbitrarily composed and shaped, weakly self-gravitating, rotating, 
deformable bodies \citep{D1,D2,D3,D4}. Such a framework yields a complete description, at 
the first post-Newtonian order, of both the global dynamics of the $N$-body system and of 
the local structure of each body by employing $N+1$ coordinate charts: one global chart 
$x^{\mu} \equiv (ct,x^i)$ $(i=1,2,3)$ covering the entire manifold $V_4$ (see below) for 
the overall dynamics of the $N$ bodies and $N$ local charts $X^{\alpha}_{\, A}\equiv 
(cT_A,X^{a}_{A})$ $(A=1,2,\dots,N)$, $(a=1,2,3)$ describing the internal dynamics of each body. 
Hereafter, inspired by Refs. \citep{D1,D2,D3,D4}, we will use the following conventions: the 
$N$ bodies are labelled by upper case Latin indices $A,B,C=1,2,\dots,N$; the global coordinates 
$x^{\mu} \equiv (ct,x^i)$ are such that spacetime indices belong to the second part of Greek 
alphabet $(\mu,\nu,\lambda,\dots)$ while purely space indices follow the second part of the 
Latin alphabet $(i,j,k,\dots)$; the local coordinate systems $X^{\alpha}_{\, A}\equiv (cT_A,X^{a}_{A})$ 
are such that spacetime indices and spatial indices are taken from the first part of the Greek 
and Latin alphabet, respectively. In order to ease the notation, sometimes the labels 
$A,B,C$ are omitted. The essential ingredient of the formalism is represented by the exponential 
parametrization (in all frames) of the metric tensor which leads to the linearization of both the 
field equations (expressible in terms of some potential functions, in strict analogy with Maxwell 
theory of electromagnetism) and the transformation laws under a change of reference system.

Let there be given a four-dimensional smooth differentiable abstract manifold $V_4$ endowed 
with $N$ abstract world lines $\mathfrak{L}_A$ $(A=1,2,\dots,N)$ and $N$ topological tubes 
$\mathcal{T}_A \subset V_4$ representing some open neighbourhood of $\mathfrak{L}_A$. 
The local chart $X^{\alpha}_{\, A}$ is said to be adapted to the world line if it maps every 
point $P$ belonging to $\mathfrak{L}_A$ onto the ``time axis'' of $\mathbb{R}^4$, i.e.,
\begin{equation}
X^{\alpha}_{\, A} : \mathcal{T}_A\rightarrow \mathbb{R}^4 \; \vert \;  X^{\alpha}_{\, A} (P)
=(S,0,0,0), \; \; P \in \mathfrak{L}_A, \; S \in \mathbb{R}.
\end{equation}
In this way it is possible to parametrize $\mathfrak{L}_A$ through the real factor $S=cT$ 
which fulfils the role of a special affine parameter having the physical dimensions of a length, 
i.e., $X^{\alpha}_{\, A} (P(S))=(S,0,0,0), \; (S \in \mathbb{R}, \; P \in \mathfrak{L}_A)$. 
Moreover, we can define a one-parameter family of vectorial bases $\boldsymbol{{\rm e}}^{A}_{\alpha}(S)$ 
along $\mathfrak{L}_A$ (i.e., a basis of the tangent vector space 
$T_{P(S)}V_4$ to $V_4$ at $P(S)\in \mathfrak{L}_A$) as
\begin{equation}
\boldsymbol{{\rm e}}^{A}_{\alpha}(S)= \left. \dfrac{\partial}
{\partial X^{\alpha}_{\, A}  } \right \vert_{P(S)},
\label{2.2a}
\end{equation}
such that 
\begin{equation}
\boldsymbol{{\rm e}}^{A}_{0}(S)= \left. \dfrac{\partial}{\partial S  } 
\right \vert_{P(S)},
\end{equation}
represents the tangent vector to the $S$-parametrized world line. 

By means of the metric-independent structure introduced above the most general coordinate 
transformation linking global and local coordinates \cite{D1}
\begin{equation}
x^\mu = f^{\mu}_{\,A}(X^{0}_{A},X^{1}_{A},X^{2}_{A},X^{3}_{A}),
\label{2.4a}
\end{equation}
can be expressed by (label $A$ omitted)
\begin{equation}
x^\mu (X^{\alpha})= z^\mu (X^0) + e_{a}^{\; \mu} (X^0)X^a + \xi^\mu (X^0,X^a),
\label{2.8aa}
\end{equation}
where (label $A$ written explicitly)
\begin{equation}
z^{\mu}_{A}(S) = f^{\mu}_{\,A}(S,0,0,0),
\end{equation}
represents the expression that the $S$-parametrized world line $\mathfrak{L}_A$ assumes in 
the global coordinate system $x^\mu$, while
\begin{equation}
e_{A\, a}^{\; \mu} (S) = \dfrac{\partial  f^{\mu}_{\,A}}{\partial X^{a}_{\, A} }(S,0,0,0),
\label{2.5ba}
\end{equation}
are the global components of the three vectors $\boldsymbol{{\rm e}}^{A}_{a}(S)$ (cf. Eq. 
(\ref{2.2a})) and eventually the term
\begin{equation}
\xi^{\mu}_{\, A} (S,X^{a}_{\,A})= f^{\mu}_{\,A} (S,X^{a}_{\, A}) - f^{\mu}_{\,A} (S,\boldsymbol{0}) 
- X^{a}_{\, A} \dfrac{\partial  f^{\mu}_{\,A}}{\partial X^{a}_{\, A} }(S,\boldsymbol{0}),
\end{equation}
is such that
\begin{equation}
\xi^\mu (X^0,X^a)= {\rm O}\left((X^a)^2\right) \; \; \; {\rm as}\;  X^a \rightarrow 0 
\; \; {\rm with \; fixed}\; X^0.
\label{2.8ba}
\end{equation}
In particular, the last condition results from the requirement that the metric admits a 
post-Newtonian expansion of the usual type (see Eq. (\ref{2.12a}) below). Note also that Eq. 
(\ref{2.5ba}) is closely connected with the values of the Jacobian matrix elements
\begin{equation}
A^{\mu}_{\, \alpha}= \dfrac{\partial x^{\mu}(X^{\beta})}{\partial X^\alpha},
\end{equation}
of the coordinate transformation (\ref{2.4a}) evaluated at the origin of the local system.
\begin{figure}
\includegraphics[scale=0.9]{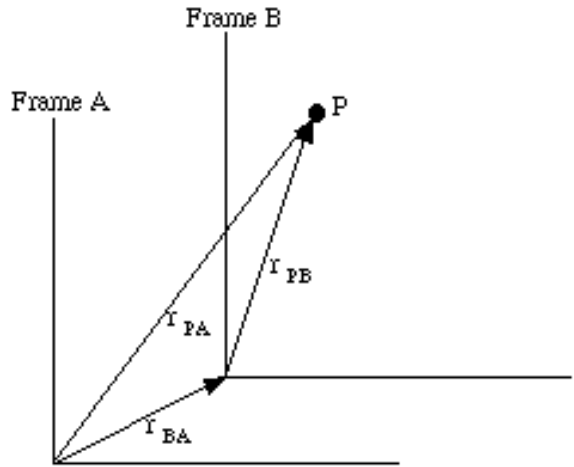}
\caption{The Newtonian decomposition of a position vector in fixed (A) and mobile (B) coordinate systems.}
\label{ClassicFrames}
\end{figure}

We believe that Eq. (\ref{2.8aa}) can be better understood once we have compared it with the Newtonian 
decomposition of a position vector $\bold{r}_{{\rm PA}}$ in two frames (Fig. \ref{ClassicFrames})
\begin{equation}
\bold{r}_{{\rm PA}}= \bold{r}_{{\rm BA}}+\bold{r}_{{\rm PB}}.
\label{NewtonDecomposition}
\end{equation}
In fact, since the term $z^\mu (X^0)$ occurring in (\ref{2.8aa}) just describes the world line 
$\mathfrak{L}_A$ in the global frame, it reminds us of the vector $\bold{r}_{{\rm BA}}$ denoting 
the position of the origin of the mobile frame as seen in the fixed one. As a result, in the 
post-Newtonian pattern the global spatial coordinates of the origin of the local coordinate system, 
following (the world line of) every body of the system, expressed as functions of the global time 
$t$ will be represented by $x^i=z^{i}(t)$. Moreover, the term $e_{a}^{\; \mu} (X^0)X^a$ appearing in 
Eq. (\ref{2.8aa}) resembles the quantity $\bold{r}_{{\rm PB}}=x_{{\rm PB}} \hat{x}_{{\rm B}}+y_{{\rm PB}} 
\hat{y}_{{\rm B}}+z_{{\rm PB}} \hat{z}_{{\rm B}}$ ($\hat{x}_{{\rm B}}$, $\hat{y}_{{\rm B}}$, 
and $\hat{z}_{{\rm B}}$ being the unit vectors of the axes of the mobile system) because, 
as pointed out before, $e_{a}^{\; \mu}$ are the global version of the vector basis of tangent 
vector space $T_{P(S)}V_4$. Finally the $\xi^{\mu}$ term of (\ref{2.8aa}) measures the deviation 
between Newtonian and post-Newtonian dynamics. Furthermore, it is essential to stress that the 
structure of Eq. (\ref{2.8aa}) results from the post-Newtonian analysis of those effacement properties 
which could allow, like in Newtonian theory, a separation of the $N$-body problem into two decoupled 
sub-problems: the internal problem concerning the motion of each body around its centre of mass and 
the external one, involving the dynamics of all centres of mass of the bodies. In fact, in the Newtonian 
framework the key point towards the achievement of such a decoupling is represented by the existence of 
some nonrotating accelerated mass-centred-frames with respect to which, in the description of the internal 
problem, the effects of the external gravitational potential acting locally on each body are suppressed, 
leaving only small (tidal) effects. The search for a good post-Newtonian definition of some 
external-gravitational-field-effacing coordinates in the analysis of the $N$-body internal problem has 
led to Eq. (\ref{2.8aa}) \cite{D1987}. This issue is strictly connected to the post-Newtonian definition 
of center-of-mass frames and will be further discussed later on.

At this stage, we can define a metric structure both in the global frame and in the $N$ local systems by 
\begin{equation}
\begin{split}
& g_{\mu \nu}(x^\lambda)= \eta_{\mu \nu} + h_{\mu \nu}(x^\lambda),\\
& G^{A}_{\; \alpha \beta} (X^\gamma)= \eta_{\alpha \beta} + H^{A}_{\; \alpha \beta} (X^\gamma), \; \; \; (A=1,2,\dots,N),
\end{split}
\end{equation}
$\eta_{\mu \nu}$ being the flat Minkowski metric, whereas $h_{\mu \nu}$ and $H^{A}_{\; \alpha \beta}$ 
represent the metric deviation in the global and in the local frames, respectively. By making the standard 
post-Newtonian assumptions for the metric ($T \equiv X^0/c$) \cite{D1,D2} 
\begin{equation}
\begin{split}
& h_{00}(t,\bold{x})= {\rm O}(c^{-2}),\;\;\;\;\;\;\;\;\; \; \; \; \; \;\;\;\;  h_{0i}(t,\bold{x})
= {\rm O}(c^{-3}),\;\;\;\;\;\;\;\;\; \;\;  \; \; \;\;\;\;h_{ij}(t,\bold{x})
= {\rm O}(c^{-2}),\;\;\;\;\;\; \; \;\;  \; \;\;\;\;\;\;\;   \\
& H^{A}_{\; 00} (T, \bold{X})= {\rm O}(c^{-2}), \;\;  \;\;\; \; \; \; \; \;\;\;\; H^{A}_{\; 0a} 
(T, \bold{X})= {\rm O}(c^{-3}),  \; \;\;\;\; \; \; \; \; \;\;\;\; H^{A}_{\; ab} (T, \bold{X})
= {\rm O}(c^{-2}), \;\;\;\;\;\;\;\;\; \; \; \; \; \;\;\;\;
\end{split}
\label{2.12a}
\end{equation}
and by postulating that the coordinate transformation (\ref{2.8aa}) involves Jacobian matrix elements 
having the form (``slow motion assumption'') \cite{D1}
\begin{equation}
A^{0}_{0}={\rm O}(c^{-0}), \; 
A^{i}_{0}={\rm O}(c^{-1}), \; 
A^{0}_{a}={\rm O}(c^{-1}), \;
A^{i}_{a}={\rm O}(c^{-0}), \;
\label{2.13a}
\end{equation}
the $z-e-\xi$ elements appearing in (\ref{2.8aa}) turn out to be highly constrained \cite{D1,D2}. 
In particular, the $3 \times 3$ time-dependent matrix $e^{i}_{a}(S)$ is such that
\begin{equation}
\begin{split}
& \delta_{ij} \,e^{i}_{a}(S) \, e^{j}_{b}(S)=e^{i}_{a}(S) \, e^{i}_{b}(S)= \delta_{ab} + {\rm O}(c^{-2}), \\
& \dfrac{{\rm d}}{{\rm d}T}e^{i}_{a}(T)={\rm O}(c^{-2}),
\end{split}
\end{equation}
resembling, modulo $\O(c^{-2})$ terms, a slowly changing Euclidean rotation matrix. The only quantity 
occurring in (\ref{2.8aa}) which is not constrained at the first post-Newtonian order by Eqs. 
(\ref{2.12a}) and (\ref{2.13a}) is represented by the term $\xi^i$, since it is found in Ref. \cite{D1} 
that $\xi^i={\rm O}(c^{-2})$. This peculiarity leaves some (gauge) freedom to the spatial coordinates 
which can be restricted by imposing in all $N+1$ frames four algebraic conditions called 
``spatial isotropy conditions'' which are represented by \citep{D1,D2,D3,D4}
\begin{equation}
\begin{split}
& -g_{00}g_{ij}= \delta_{ij} + {\rm O}(c^{-4}), \\
& -G^{A}_{\; 00} G^{A}_{\; ab} = \delta_{ab}+  {\rm O}(c^{-4}), \;\;\;\;\;\;\,\; \forall\;A.
\label{2.29a}
\end{split}
\end{equation}
The coordinates selected by Eq. (\ref{2.29a}) are referred to as ``conformally Cartesian'' or 
``isotropic''. The gauge choice (\ref{2.29a}) fixes completely the spatial coordinate freedom in all 
frames up to time-dependent isometries of Euclidean three-space. In other words, Eq. (\ref{2.29a}) 
leaves a gauge freedom in the time coordinate at ${\rm O}(c^{-4})$ in all coordinate systems. 
Moreover, the spatial isotropy conditions (\ref{2.29a}) imply that the three spatial coordinates 
$x^i$ are harmonic at the second post-Newtonian order, i.e.,
\begin{equation}
\square_{g}x^i= {\rm O}(c^{-4}),
\label{3.15a}
\end{equation}
$\square_{g} = \left(1/\sqrt{-g}\right) \partial_{\mu} \left(\sqrt{-g} \, g^{\mu \nu} \partial_{\nu}\right)$ 
being the d' Alembert wave operator acting on scalar functions. Since Eq. (\ref{3.15a}) follows directly 
from the imposition of the (spatial part of the full) standard harmonic and post-Newtonian gauges, 
we see that (\ref{2.29a}) encompasses both these choices. 

At this stage, by enforcing both the assumptions (\ref{2.12a}) and (\ref{2.13a}) and the spatial isotropy 
conditions (\ref{2.29a}) the structure of the transformation (\ref{2.8aa}) is further constrained, leaving 
only the usual Newtonian freedom involving the choice of an arbitrarily moving origin $z^i(t)$ of the local 
frame (we will see that a preferred choice is the one for which the Blanchet-Damour mass dipole 
vanishes \cite{BD}) and of a slowly changing SO$(3)$ rotation matrix $R^{i}_{a}(T)$ describing the 
rotational state of the local spatial coordinate grid (together with the gauge freedom for time 
coordinate we have just mentioned). The matrix $R^{i}_{a}(T)$ is defined through the $3\times3$ 
matrix $e^{i}_{a}(T)$, since it turns out to be proportional, apart from second-order terms, to 
the space-space part of a general Lorentz transformation (i.e., a boost combined with an 
arbitrary rotation matrix) \cite{D1}
\begin{equation}
\bigg(1+ {\rm O}\left(c^{-2}\right)\bigg)e^{i}_{a}(T)= \left( 1+ \dfrac{1}{2c^2} \bold{v}^2 \right) 
\left(\delta^{ij} + \dfrac{1}{2c^2} v^i v^j \right) R^{j}_{a}(T) + {\rm O}(c^{-4}),
\end{equation}
$b^{ij}=\delta^{ij}+\dfrac{v^{i} v^{j}}{2c^2}$ being the boost components with 
\begin{equation}
v^i = \dfrac{{\rm d}z^i}{{\rm d}t}= \dfrac{{\rm d}z^i}{{\rm d}T}+ {\rm O}(c^{-2}),
\end{equation}
representing the velocity of the origin of the local frame. The slow post-Newtonian 
evolution of $R^{i}_{a}(T)$ is caught by
\begin{equation}
\dfrac{{\rm d}}{{\rm d}T} R^{i}_{a}(T)={\rm O}(c^{-2}),
\end{equation}
while its SO$(3)$ structure is enlightened through the relations
\begin{equation}
\begin{split}
& R^{i}_{a}(T) \, R^{j}_{a}(T) = \delta^{ij}, \\
& R^{i}_{a}(T) \, R^{i}_{b}(T) = \delta_{ab}.
\end{split}
\end{equation} 
At this stage, we can introduce the exponential parametrization of the ten independent unknown 
metric tensor components through the potential $w_{\mu}=(w,w_{i})$, starting from 
the global frame \cite{D1,BD}
\begin{equation}
\begin{split}
& g_{00}= -\E^{-2w/c^2}, \\
& g_{0i}=-4w_{i}/c^3,\\
& g_{ij}=\gamma_{ij} \, \E^{2w/c^2}, 
\label{3.3a}
\end{split}
\end{equation}
$\gamma_{ij}$ being the three-metric associated to $g_{\mu \nu}$ by
\begin{equation}
\gamma_{ij}=-g_{00}g_{ij}+g_{0i}g_{0j}.
\end{equation}
By exploiting Eq. (\ref{3.3a}) and the post-Newtonian assumptions (\ref{2.12a}), six of the 
Einstein equations imply that the spatial coordinates are Cartesian coordinates for the three-metric, i.e., 
\begin{equation}
\gamma_{ij}=\delta_{ij}+ \O(c^{-4}),
\label{3.6a}
\end{equation}
whereas the four remaining Einstein equations give coupled {\it linear} partial differential equations 
for $w$ and the three-vector components $w^{i}=w_{i}$ 
expressed by ($\partial_t=c\partial_0=\partial/\partial t$, 
$w^i \equiv \gamma^{ij}w_j$) \cite{D1} 
\begin{equation}
\begin{split}
& \square w + \dfrac{4}{c^2} \partial_t \left(\partial_t w + \partial_i w_i \right)
= - 4 \pi G \sigma + \O(c^{-4}), \\
&  \triangle w_i - \partial_{i}\partial_{j}w_{j} - \partial_t \partial_i w
= - 4 \pi G \sigma_i + \O(c^{-2}), 
\label{3.11a}
\end{split}
\end{equation}
where $- \triangle$ is the flat-space Laplacian (in general we would have $\triangle = \gamma^{ij} D_j D_i$, 
$D_i$ being the (spatial) covariant derivative associated to $\gamma_{ij}$), while $\sigma$ and $\sigma_i$ 
represent a gravitational mass density and a mass current density, respectively. They are defined in 
terms of the contravariant components of the (post-Newtonian) stress-energy tensor\footnote{$T^{\mu \nu}$ 
satisfies the usual post-Newtonian assumptions for the matter:
\begin{equation}
\begin{split}
& T^{00}= \O(c^2), \\
& T^{0i}=\O(c), \\
& T^{ij}=\O(c^0).
\label{2.25a}
\end{split}
\end{equation}} as
\begin{equation}
\begin{split}
& \sigma \equiv \dfrac{T^{00}+T^{jj}}{c^2}, \\
& \sigma^i \equiv \dfrac{T^{0i}}{c}. 
\end{split}
\end{equation}
Note that the definitions given above are independent of the particular form taken by $T^{\mu \nu}$, 
which is supposed to have a very {\it general} structure, being subjected only to conditions (\ref{2.25a}). 
Equations (\ref{3.11a}) are found to be gauge invariant, apart from post-Newtonian error terms, 
under the transformations \cite{D1}
\begin{equation}
\begin{split}
& w \rightarrow w^\prime= w - \dfrac{1}{c^2} \partial_t \lambda(x^\mu), \\
& w_i \rightarrow w_{i}^{\prime}= w_i + \dfrac{1}{4} \partial_i \lambda(x^\mu), 
\label{3.12a}
\end{split}
\end{equation} 
$\lambda(x^\mu)$ being an arbitrary differentiable function. The (approximate) gauge invariance 
(\ref{3.12a}) shows a manifest analogy with typical $U(1)$ transformations underlying Maxwell theory and it 
is related to the above-mentioned time coordinate freedom, being expressible as a shift
\begin{equation}
\delta t = c^{-4} \lambda(x^\mu).
\end{equation}
If we further exploit the similarity with electromagnetic theory, we can define the gravitational field strength as
\begin{equation}
b_{\mu \nu} = \partial_\mu a_{\nu} - \partial_\nu a_{\mu},
\label{3.21aa}
\end{equation}
with $a_{\mu} \equiv (cw,-4w_i)$ and hence the (global) gauge-invariant gravitoelectric and 
gravitomagnetic fields \cite{D1}
\begin{equation}
\begin{split}
& e_i(w)= \partial_i w + \dfrac{4}{c^2} \partial_t w_i, \\
& b_{ij}(w) = \epsilon_{ijk} b_k (w)= -4 (\partial_i w_j - \partial_j w_i ),
\label{3.21bca}
\end{split}
\end{equation}
respectively. 

Bearing in mind that (\ref{2.29a}) implies that $x^i$ coordinates are in the kernel of the wave operator 
(up to $\O(c^{-4})$ terms, cf. Eq. (\ref{3.15a})), if we further constrain the (gauge) time variable 
to be harmonic through the condition
\begin{equation}
\square_g x^0 = -\dfrac{4}{c^3} \left( \partial_t w + \partial_i w_i \right) + \O(c^{-5})=0, 
\end{equation}
Eq. (\ref{3.11a}) reduces to the field equations
\begin{equation}
\begin{split}
& \square w = \triangle w -\dfrac{1}{c^2} \partial^{2}_{t} w = -4 \pi G \sigma + \O(c^{-4}), \\
& \triangle w_i = -4 \pi G \sigma_i + \O(c^{-2}). 
\label{3.18a}
\end{split}
\end{equation}

The linearity of Eq. (\ref{3.11a}) makes it possible to express its general solution 
$w^{{\rm general}}_{\mu}$ as
\begin{equation}
w^{{\rm general}}_{\mu}=w^{N}_{\mu}+{\bar w}^{N}_{\mu},
\label{3.24aa}
\end{equation}
${\bar w}^{N}_{\mu}$ representing the general solution of the homogeneous system associated 
to (\ref{3.11a}), i.e.,
\begin{equation}
\mathcal{L}^{\mu}[{\bar w}^{N}_{\nu}]=0,
\end{equation}
$w^{N}_{\mu}$ being a particular solution of the inhomogeneous system (\ref{3.11a}), written as
\begin{equation}
\mathcal{L}^{\mu}[w^{N}_{\nu}]=-4 \pi G \sum_{A=1}^{N} \sigma^{\mu}_{A},
\end{equation}
where $\sigma^{\mu}_{A}$ denotes the source contribution of each body of the system. For an isolated 
$N$-body system we can always set ${\bar w}^{N}_{\mu}=0$. Therefore, the potential for 
the global $N$-body metric reads as
\begin{equation}
w^{N}_{\mu}=\sum_{A=1}^{N} w_{\mu}^{A}.
\label{3.29aa}
\end{equation}
If we employ the so-called harmonic gauge, the error occurring in the field equations 
(\ref{3.18a}) satisfied by $w_i$ allows us to write
\begin{equation}
\square = \triangle + \O(c^{-2}),
\end{equation}
and hence the contributions, generated by each body, to the global $N$-body metric appearing 
in Eq. (\ref{3.29aa}) can be expressed as
\begin{equation}
w^{\mu}_{A}(x^\lambda)= \square^{-1}_{\rm sym}(-4 \pi G \sigma^{\mu}_{A}),
\label{3.29ba}
\end{equation}
$\square^{-1}_{\rm sym}$ denoting the half sum of the retarded and advanced flat-space Green's 
functions. Of course, strictly speaking, the inverse of the wave operator is an integral operator 
with kernel given by the Green functions we have mentioned.

As long as (\ref{3.6a}) (or equivalently (\ref{2.29a})) holds, the form of both gravitoelectric 
and gravitomagnetic fields (\ref{3.21bca}) will be the same in any coordinate system and hence a 
linear gauge invariant description of the gravitational field will be always possible, at least 
to first post-Newtonian order. Therefore, in each local frame $X^{\alpha}_{\, A}$ the local metric 
$G^{A}_{\; \alpha \beta}(X^{\gamma}_{\, A})$ satisfying the isotropy condition (\ref{2.29a}) can be 
exponentially parametrized through the local potential $W^{A}_{\alpha}=(W, W^{A}_{a})$ defining, 
by the same Eqs. (\ref{3.21aa}) and (\ref{3.21bca}), the local gravitoelectric and gravitomagnetic 
fields $E^{A}_{a}$ and $B^{A}_{a}$, i.e., (label $A$ omitted) \cite{D1}
\begin{equation}
\begin{split}
& G_{00}= -\E^{-2W/c^2}, \\
& G_{0a}=-4W_{a}/c^3,\\
& g_{ij}=\gamma_{ab} \, \E^{2W/c^2}, 
\label{4.1a}
\end{split}
\end{equation}
\begin{equation}
\begin{split}
& E_a(W)= \partial_a W + \dfrac{4}{c^2} \partial_T W_a, \\
& B_{ab}(W) = \epsilon_{abc} B_c (W)= -4 (\partial_a W_b - \partial_b W_a),
\label{4.14a}
\end{split}
\end{equation}
the local potential $W^{A}_{\alpha}$ satisfying Eqs. (\ref{3.11a}), 
which in $X^{\alpha}_{\, A}$ coordinates read as
\begin{equation}
\begin{split}
& \square_X W^A + \dfrac{4}{c^2} \partial_T \left(\partial_T W^A + \partial_a W_{a}^{A} \right)
= - 4 \pi G \Sigma^A + \O(c^{-4}), \\
&  \triangle_X W^{A}_{a} - \partial_{a}\partial_{T}W^{A} - \partial_a \partial_b W^{A}_{b}
= - 4 \pi G \Sigma_{a}^{A} + \O(c^{-2}), 
\label{4.3a}
\end{split}
\end{equation}
where $\square_X=\triangle_X-\dfrac{1}{c^2}\partial^{2}_{T}$, $\triangle_X= \partial^a \partial_a$ and
\begin{equation}
\begin{split}
& \Sigma_A \equiv \dfrac{T^{\,00}_{A}+T^{\,aa}_{A}}{c^2}, \\
& \Sigma^{a}_{A} \equiv \dfrac{T^{\,0a}_{A}}{c},
\end{split}
\label{4.4a}
\end{equation}
the only non-vanishing components of the stress-energy tensor, now defined in the local frame 
$X^{\alpha}_{\, A}$, being associated to the body $A$ itself. By exploiting the linearity of (\ref{4.3a}), 
the solution $W^{A}_{\alpha}$ reads as
\begin{equation}
W^{A}_{\alpha} = W^{+\,A}_{\alpha} + \overline{W}^{A}_{\alpha},
\label{4.7a}
\end{equation}
where $ W^{+\,A}_{\alpha}$ denotes the locally generated part of the potential, i.e., a particular 
solution of the inhomogeneous system (\ref{4.3a}) being solved in the harmonic gauge
\begin{equation}
W^{+\,A}_{\alpha} = \square_{X,{\rm sym}}^{-1} \left(- 4 \pi G \Sigma_{\alpha}^{A}\right),
\label{4.5a}
\end{equation}
while $\overline{W}^{A}_{\alpha}$ represents the external part of $W^{A}_{\alpha}$, since it 
satisfies, in the domain of the local chart $X^{\alpha}_{\, A}$ (i.e., the domain containing the 
body $A$ and no other bodies $B \neq A$), the homogeneous system associated to (\ref{4.3a}). 

We have already stated that the investigation of the Newtonian $N$-body internal problem is carried 
out by employing accelerated centre-of-mass frames having local 
coordinates (cf. Eq. (\ref{NewtonDecomposition})) 
\begin{equation}
X^{i}_{A} = x^i - z^{i}_{A} (t), \;\;\; \; \; \;\;\;\;\; \; \;\;\;\;(i=1,2,3),
\label{1.2a}
\end{equation}
($z^{i}_{A}$ denoting the global coordinates of the barycentre of the body $A$) with respect to which 
the external potential gets replaced by the effective gravitational potential (whose gradient governs the 
motion of the mass elements in each barycentric frame)
\begin{equation}
U^{{\rm (eff)}}_{A}(\bold{X}_A)= U^{{\rm (ext)}}_{A}(\bold{X}_A+\bold{z}_A) - U^{{\rm (ext)}}_{A}
(\bold{z}_A)-\dfrac{{\rm d^2}\bold{z}_A}{{\rm d}t^2} \cdot \bold{X}_A,
\label{6.3a}
\end{equation}
the last term being (fictitious) inertial forces. 

By employing Eq. (\ref{6.3a}) from a relativistic point of view (i.e., the potential $U$ is replaced 
by the metric tensor, six components of which are gauged away by Eq. (\ref{2.29a})), it is possible to 
show that the local and global gravitational potentials (i.e., the four metric tensor components 
``surviving'' Eq. (\ref{2.29a})) are related by a transformation law having the form (label $A$ omitted) \cite{D1}
\begin{equation}
w_\mu (x^\nu)= \mathcal{A}_{\mu \alpha}(T) W^{\alpha}(X^\gamma) + \mathcal{B}_\mu (X^\gamma),
\label{4.12da}
\end{equation} 
the inverse being given by
\begin{equation}
W_\alpha(X^\gamma)= ( \mathcal{A}^{-1})^{\mu}_{\,\alpha}(T) 
\left(w_\mu (x^\nu)- \mathcal{B}_\mu (X^\gamma)\right),
\label{4.13aa}
\end{equation}
$X^\gamma$ and $x^\nu$ denoting, respectively, the local and the global coordinates of the same 
spacetime event and where the coefficients $\mathcal{A}_{\mu \alpha}(T)$ and $\mathcal{B}_\mu (X^\gamma)$ 
depend on global and local components of the velocity $v^{i}={\rm d}z^i/{\rm d}t$ of the origin of the 
local frame and on the components of the Jacobian matrix of (\ref{2.8aa}). The inertial terms occurring in 
Eq. (\ref{6.3a}) are represented by the coefficients $\mathcal{B}_\mu$, so that the gauge invariant 
local fields defined by Eq. (\ref{4.14a}) can be split as
\begin{equation}
\begin{split}
& \bold{E}= \bold{E}^{\prime} + \bold{E}^{\prime \prime}, \\
& \bold{B}= \bold{B}^{\prime} + \bold{B}^{\prime \prime}, \\
\end{split}
\end{equation}
with
\begin{equation}
\begin{split}
& \bold{E}^{\prime} \equiv \bold{E}^{\prime} \left[ \mathcal{A}_{\mu \alpha}^{-1}w^\mu\right], 
\; \; \; \;\; \; \; \;\; \; \; \; \bold{E}^{\prime \prime} \equiv \bold{E}^{\prime \prime} 
\left[ -\mathcal{A}_{\mu \alpha}^{-1}\mathcal{B}^\mu\right], \\
& \bold{B}^{\prime} \equiv \bold{B}^{\prime} \left[ \mathcal{A}_{\mu \alpha}^{-1}w^\mu\right],
\; \; \; \; \; \; \; \;\; \; \; \; \bold{B}^{\prime \prime} \equiv \bold{B}^{\prime \prime} 
\left[ -\mathcal{A}_{\mu \alpha}^{-1}\mathcal{B}^\mu\right], 
\end{split}
\end{equation}
the terms $\bold{E}^{\prime \prime}$ and $\bold{B}^{\prime \prime}$ being referred to as inertial 
fields. Moreover, from Eq. (\ref{4.12da}) if follows that the locally generated potential 
$W^{+\,A}_{\alpha}$ defined (in the harmonic gauge) by Eqs. (\ref{4.7a}) and (\ref{4.5a})  
is linked to the $A$-part of the local potential $w_{\mu}^{A}$ (cf. Eqs. (\ref{3.24aa}) and 
(\ref{3.29aa})) by the homogeneous transformation \cite{D1}
\begin{equation}
\begin{split}
& w_{0}^{A}(x^\nu)= \mathcal{A}^{A}_{\, 0 \alpha}(X^0)W^{+\,A}_{\alpha}(X^\beta) + \O(c^{-4}), \\
& w_{i}^{A}(x^\nu)= \mathcal{A}^{A}_{\, i \alpha}(X^0)W^{+\,A}_{\alpha}(X^\beta) + \O(c^{-2}), \\
\end{split}
\end{equation}
whereas the external local potential $\overline{W}^{A}_{\alpha}$ is related to the part of the global 
potential generated by all bodies $B \neq A$ through the inhomogeneous relation
\begin{equation}
\begin{split}
& \sum_{B \neq A} w_{0}^{B}(x^\nu)= \mathcal{A}^{A}_{\, 0 \alpha}(X^0) \overline{W}^{A}_{\alpha}(X^\beta) 
+ \mathcal{B}_{0}^A (X^\beta) + \O(c^{-4}),\\
& \sum_{B \neq A} w_{i}^{B}(x^\nu)= \mathcal{A}^{A}_{\, i \alpha}(X^0) \overline{W}^{A}_{\alpha}(X^\beta) 
+ \mathcal{B}_{i}^A (X^\beta) + \O(c^{-2}),\\
\end{split}
\label{4.55a}
\end{equation}
$\mathcal{A}^{A}_{\, \mu \alpha}$ and $\mathcal{B}_{\mu}^{A}$ being the $A$-part 
of the coefficients appearing in (\ref{4.12da}). 

\section{Newtonian tidal and multipole expansions} 

In order to deal with the $N$-body problem in Newtonian gravity, the whole problem is separated, 
as we have anticipated before, into two sub-problems, i.e., the external problem, underlying the dynamics 
of the $N$ centres of mass and the internal problem, concerning the motion of each body about its 
centre of mass \cite{D1987,Fock1959}. We introduce a set of Cartesian coordinates $x^i$ whose indices 
are raised and lowered through the Euclidean metric $\delta_{ij}$ so that $A^i=A_i$, and repeated
indices will be summed over.

We suppose that the system is isolated and the bodies are widely separated and finite in extension so that 
the dimensionless coupling measuring the force experienced by the bodies is such that
\begin{equation}
\alpha \equiv \dfrac{L}{R} \ll 1,
\label{28-1987}
\end{equation}
$L$ being their characteristic linear dimension and $R$ their separation. Moreover, the internal 
structure of bodies is governed by the isentropic equation of state of a perfect fluid where the 
pressure $p$ is a function of the mass density $\rho$, i.e., 
\begin{equation}
p=p(\rho).
\end{equation}
The local fluid motion is governed by the Euler equations
\begin{equation}
\rho \left(\dfrac{\partial}{\partial t} + v^j \dfrac{\partial}{\partial x^j} \right)v_{i}
=- \dfrac{\partial p}{\partial x^i}+ \rho \dfrac{\partial U}{\partial x^i},
\label{3-1987}
\end{equation}
where $v(t,\bold{x})$ is the fluid velocity field and the gravitational potential $U(t,x^i)$ 
of the system is represented by the solution of the Poisson equation
\begin{equation}
\triangle U = - 4 \pi G \rho,
\label{Poisson1}
\end{equation} 
subjected to the boundary condition that it should be bounded everywhere and vanish at infinity
\begin{equation}
\lim_{\substack{\vert \bold{x}\vert \rightarrow \infty \\ t={\rm const.}}} U(t,\bold{x})=0,
\label{Poisson2}
\end{equation}
expressing mathematically the physical hypothesis of having considered an isolated system. The 
unique solution of Eqs. (\ref{Poisson1}) and (\ref{Poisson2}) reads as
\begin{equation}
U(t,\bold{x}) = G \int \dfrac{\rho(t,\bold{x}^{\prime})}{\vert \bold{x}-\bold{x}^{\prime}\vert} 
{\rm d}^3x^{\prime},
\label{5-1987}
\end{equation}
$\vert \bold{x}-\bold{x}^{\prime}\vert$ denoting the Euclidean distance between the field point 
$\bold{x}$ and the source point $\bold{x}^{\prime}$, and ${\rm d}^3x^{\prime}$ being the Euclidean 
volume element. Furthermore, it is a well-known fact that the centre of mass theorem implies that 
the dynamics of the system is governed by the set of $N$ differential equations
\begin{equation}
M_A \dfrac{{\rm d}^2 z^{i}_A}{{\rm d}t^2}= \int_{V_A} \mathcal{F}^i \, {\rm d}^3x, 
\;\;\;\;\;\;\;\;\;\; (A=1,2,\dots,N),
\label{11-1987}
\end{equation} 
where $M_A=\int_{V_A} \rho(t,\bold{x}){\rm d}^3x$ is the total mass of the $A$th body occupying 
the volume $V_A$, $z^{i}_A$ denotes the components of its centre of mass, defined classically as
\begin{equation}
z^{i}_A(t) = \dfrac{1}{M_A} \int_{V_A}x^i \rho(t,\bold{x}){\rm d}^3x,
\label{7-1987}
\end{equation}
and $\mathcal{F}^i$ represents the local force density which, in a perfect fluid model, assumes 
from (\ref{3-1987}) the form
\begin{equation}
\mathcal{F}^i = \mathcal{F}_{i}= 
- \dfrac{\partial p}{\partial x^i}+ \rho \dfrac{\partial U}{\partial x^i}.
\end{equation} 
The force density can be decomposed, {\it within each body}, into an internal force (or self force)
\begin{equation}
\mathcal{F}^{ \, ({\rm int})}_{i \; A}= - \dfrac{\partial p}{\partial x^i}
+ \rho \dfrac{\partial U^{({\rm int})}_A}{\partial x^i},
\end{equation} 
the self part $ U^{({\rm int})}_A$ of the gravitational potential (\ref{5-1987}) being defined by
\begin{equation}
U^{({\rm int})}_A(t,\bold{x}) \equiv G\int_{V_A} \dfrac{\rho(t,\bold{x}^{\prime})}
{\vert \bold{x}-\bold{x}^{\prime}\vert} {\rm d}^3x^{\prime},
\end{equation}
and an external force
\begin{equation}
\mathcal{F}_{i \; A}^{({\rm ext})}=  \rho \dfrac{\partial U^{({\rm ext})}_A}{\partial x^i},
\label{14-1987}
\end{equation}
with
\begin{equation}
U^{({\rm ext})}_A(t,\bold{x}) \equiv \sum_{B \neq A}G\int_{V_B} \dfrac{\rho(t,\bold{x}^{\prime})}
{\vert \bold{x}-\bold{x}^{\prime}\vert} {\rm d}^3x^{\prime}.
\label{16-1987}
\end{equation}
By exploiting, as pointed out before, the local coordinates (\ref{1.2a}) in order to deal with 
the internal problem, the dynamics underlying the motion in the centre of mass frame of the $A$th 
body is governed by (cf. Eq. (\ref{3-1987}))
\begin{equation}
\rho \left(\dfrac{\partial }{\partial t} + u^j_{A}\dfrac{\partial }{\partial X^j_{A}} \right)u^i_{A}
=\mathcal{F}^{i \, ({\rm int})}_{A}+ \mathcal{F}^{i \, 
({\rm ext})}_{A}-\rho \dfrac{{\rm d}^2z^i_{A}}{{\rm d}t^2},
\label{20-1987}
\end{equation}
$u^i_A={\rm d}X^i_{A}/{\rm d}t=v^i - {\rm d}z^i_{A}/{\rm d}t $ being the classical relative velocity 
of a material point of the body $A$ and where we have exploited the fact that its directional derivative 
in the direction defined by $\dot{\vec{z}}_A$ ($\dot{z}^j_A={\rm d}z^j_A/{\rm d}t$) is such that
\begin{equation}
\dot{z}^j_A \dfrac{\partial u^i_A}{\partial X^j_A} = \dot{\vec{z}}_A \cdot \vec{\nabla}_{\vec{X}_A}u^i_A=0,
\end{equation} 
since $\dot{\vec{z}}_A$ does not define any direction in the centre-of-mass frame, where instead $u^i_A$ 
lives. On the other hand, the external problem is described by (see Eq. (\ref{11-1987})) 
\begin{equation}
M_A \dfrac{{\rm d}^2 z_{i \; A}}{{\rm d}t^2}= \int_{V_A} 
\mathcal{F}^{ \, ( {\rm ext})}_{i \; A}{\rm d}^3x 
+ \int_{V_A} \mathcal{F}^{ \, ({\rm int})}_{i \; A}{\rm d}^{3}X_{A}.
\label{19-1987}
\end{equation}
At this stage, let us pay more attention to the internal problem (\ref{20-1987}). Bearing in mind 
Eqs. (\ref{1.2a}), (\ref{14-1987}), and (\ref{16-1987}), it follows that
\begin{equation}
\begin{split}
\mathcal{F}^{ \, ({\rm ext})}_{i \; A} & =  \rho \dfrac{\partial U^{({\rm ext})}_A(t,\bold{x})}{\partial x^i}
= \rho \dfrac{\partial }{\partial x^i} U^{({\rm ext})}_A(t,\bold{X}_A+\bold{z}_A(t)) 
= \rho \dfrac{\partial}{\partial X^j_A}U^{({\rm ext})}_A(t,\bold{X}_A+\bold{z}_A(t))
\dfrac{\partial X^j_A}{\partial x^i} \\
& = \rho \dfrac{\partial}{\partial X^i_A}U^{({\rm ext})}_A(t,\bold{X}_A+\bold{z}_A(t)),
\end {split}
\end{equation}
and hence the last two terms on the right-hand side of (\ref{20-1987}) become
\begin{equation}
\begin{split}
& \mathcal{F}^{ \, ({\rm ext})}_{i \; A}- \rho \dfrac{{\rm d}^{2}z_{i \; A}}{{\rm d}t^2} 
= \rho \left( \dfrac{\partial}{\partial X^i_A}U^{({\rm ext})}_A(t,\bold{X}_A+\bold{z}_A(t)) 
- \ddot{ z}^i_A \right) \\
& = \rho \dfrac{\partial}{\partial X^i_A} \left( U^{({\rm ext})}_A(t,\bold{X}_A+\bold{z}_A(t)) 
- \ddot{ z}^j_A X^j_A \right) \\
&= \rho \dfrac{\partial}{\partial X^i_A} \left( U^{({\rm ext})}_A(t,\bold{X}_A+\bold{z}_A(t)) 
-C(t) - \ddot{ z}^j_A X^j_A \right),
\end{split}
\end{equation}
$C(t)$ being a generic differentiable function of time. The last term of the above equation represents 
the most general definition of effective potential. On choosing, for the sake of simplicity, 
\begin{equation}
C(t) \equiv U^{({\rm ext})}_A(\bold{z}_A(t)),
\end{equation}
we recover the definition (\ref{6.3a}) of effective potential and hence the internal problem 
(\ref{20-1987}) reads as
\begin{equation}
\rho \left(\dfrac{\partial }{\partial t} + u^j_{A}\dfrac{\partial }{\partial X^j_{A}} \right)u_{i \; A}
=\mathcal{F}^{ \, ({\rm int})}_{i \; A}+ \rho\dfrac{\partial}{\partial X^i_A} U^{\rm (eff)}_A(\bold{X}_A).
\label{internal_problem}
\end{equation}
It is now clear that both the external problem (\ref{19-1987}) and the internal one (\ref{internal_problem}) 
are strongly coupled. In fact, in Eq. (\ref{19-1987}) the first term on the right-hand side has got both 
internal and external nature, while the second term is purely internal. Moreover, the internal problem 
(\ref{internal_problem}) is characterized by the presence of the effective potential, which clearly 
shows an external structure. However, the choice of the local (external-field-effacing) barycentric 
coordinates (\ref{1.2a}) and the occurrence of some physical effects lead to the effacement of these 
``spurious'' terms, allowing a clear separation of the two problems. We briefly mention 
these issues \cite{D1987,Fock1959}.

As far as the external problem is concerned, the second term on the right-hand side of (\ref{19-1987}), 
being the total self force acting on $A$, is clearly forced to vanish because of the third Newton's 
law of motion. Therefore, the external problem is now governed by
\begin{equation}
M_A \dfrac{{\rm d}^2 z_{i \; A}}{{\rm d}t^2}= \int_{V_A} \rho(t,\bold{X}_A+\bold{z}_A)\dfrac{\partial}
{\partial x^i}U^{({\rm ext})}_A(t,\bold{z}_A+\bold{X}_A){\rm d}^3X_A.
\label{29-1987}
\end{equation}
We will shortly see that the hypothesis according to which the bodies of the system are widely 
separated (cf. Eq. (\ref{28-1987})) is crucial in proving the effacement of the internal structure 
from the external problem. In fact, we will introduce two simultaneous Taylor expansions. First of 
all, we Taylor expand $\dfrac{\partial}{\partial x^i}U^{({\rm ext})}_A(\bold{z}_A+\bold{X}_A) 
\equiv U^{({\rm ext})}_A(\bold{z}_A+\bold{X}_A)_{,i}$ about $\bold{X}_A=\bold{0}$, i.e., a power 
series in $\bold{X}_A=\bold{0}$ called tidal expansion, according to
\begin{equation}
U^{({\rm ext})}_A(\bold{z}_A+\bold{X}_A)_{,i}=U^{({\rm ext})}_A(\bold{z}_A)_{,i} 
+ \dfrac{1}{2}U^{({\rm ext})}_A(\bold{z}_A)_{,ijk}X^{j}_A X^k_A + \O (\vert\bold{X}_A \vert^3 ),
\label{30-1987}
\end{equation}
giving, jointly with (\ref{29-1987}), the expansion in $\alpha$
\begin{equation}
M_A \dfrac{{\rm d}^2 z_{i \; A}}{{\rm d}t^2} = M_A U^{({\rm ext})}_A(\bold{z}_A)_{,i} 
+ \dfrac{1}{2} I^{jk}_A U^{({\rm ext})}_A(\bold{z}_A)_{,ijk} + \O(\alpha^3),
\label{31-1987}
\end{equation}
where we have defined the mass moment of the body $A$ as
\begin{equation}
I^{ij \dots k}_A(t) \equiv \int_{V_A} \rho(t,\bold{z}_A+\bold{X}_A) X^i_A X^j_A\dots X^k_A  \, {\rm d}^3X_A.
\label{Newtonian-Mass-Moment}
\end{equation}
In particular, the symmetric tensor
\begin{equation}
I^{ij}_A(t)=\int_{V_A} \rho(t,\bold{z}_A+\bold{X}_A) X^i_A X^j_A \, {\rm d}^3X_A = I^{ji}_A(t),
\label{33-1987}
\end{equation}
denotes the second-order relative mass moment and 
\begin{equation}
I^{i}_A(t)=\int_{V_A} \rho(t,\bold{z}_A+\bold{X}_A) X^i_A \, {\rm d}^3X_A=0,
\label{32-1987}
\end{equation}
since the $X^i_A$ frame is such that its origin coincides with the centre of mass of $A$ (see the 
Newtonian definition (\ref{7-1987})). Moreover, we have also exploited the fact that, on dimensional 
ground, ($\rho(t,\bold{z}_A+\bold{X}_A) \equiv \rho_A(t,\bold{X}_A)$ being the mass density of $A$ and the 
square brackets denoting the physical dimension of the derivatives of the potential)
\begin{equation}
\int_{V_A} {\rm d}^3X_A \,\rho_A(t,\bold{X}_A)\, \O (\vert\bold{X}_A \vert^3 ) \sim M_A\, L^3 \,
\left[U^{({\rm ext})}_{A\,,ijkl}\right] \sim \alpha^3 M_A \left[U^{({\rm ext})}_{A\,,i}\right],
\footnote{Recall from Eq. (\ref{31-1987}) that $\O (\vert\bold{X}_A \vert^3 )$ 
contains implicitly spatial indices.}
\label{alpha^3}
\end{equation}
i.e., $\alpha^3$ smaller than the first term on the right-hand side of (\ref{31-1987}). Furthermore, 
bearing in mind that the functions $\rho(t,\bold{x}^{\prime})$ occurring on the right-hand side of Eq. 
(\ref{16-1987}) refer to mass densities of the bodies $B \neq A$ (i.e, the source terms of the external 
potential), it follows that the Laplacian of $U^{({\rm ext})}_A(\bold{X}_A+\bold{z}_A)$ vanishes when 
evaluated within the body $A$. As a result, the trace part of the mass moments appearing in (\ref{31-1987}) 
give no contribution. In addition, the Schwarz theorem about mixed partial derivatives allows us to get rid 
also of the antisymmetric part of (\ref{Newtonian-Mass-Moment}) at any order in (\ref{31-1987}). Thus, 
we define the Newtonian multipole moment $Q^{ij \dots k}_A(t)$ of the body $A$ as the symmetric trace-free 
part of $I^{ij \dots k}_A(t)$, which we denote by enclosing the indices within the symbol $\langle \cdot \rangle $, i.e., 
\begin{equation}
Q^{ij \dots k}_A(t) \equiv   I^{\langle ij \dots k \rangle}_A(t) .
\label{Newtonian-Multipole-moments}
\end{equation}
For instance, bearing in mind that for a generic rank-two tensor $T^{ij}$ we have 
\begin{equation}
T^{\langle ij \rangle}    = T^{(ij)} - \dfrac{1}{3} \delta^{ij} T^{kk},
\end{equation} 
$T^{(ij)} $ denoting the symmetric part of $T^{ij}$, it easily follows that
\begin{equation}
Q^{ij}_A(t) =I^{ij}_A(t) - \dfrac{1}{3} \delta^{ij} I^{kk}_A(t),
\end{equation}
once the symmetry property showed up in Eq. (\ref{33-1987}) has been exploited. $Q^{ij}_A(t)$ is 
referred to as quadrupole moment of $A$ \cite{D1987}. The centre-of-mass-frame condition can now be expressed as 
the vanishing of the dipole moment (cf. Eq. (\ref{32-1987})), i.e.,
\begin{equation}
Q^i_A(t)=0,
\label{Newtonian-Mass-Centred-Frame}
\end{equation}
whereas the total mass $M_A$ (see below Eq. (\ref{11-1987})) can be conceived as the monopole moment 
of the body $A$. We will see that within the relativistic framework the Newtonian multipole moments 
(\ref{Newtonian-Multipole-moments}) will be replaced by Blanchet-Damour 
multipole moments \cite{BD,D1}. However, 
at this stage we have realized that Eq. (\ref{31-1987}) is equivalent to
\begin{equation}
M_A \dfrac{{\rm d}^2 z_{i \; A}}{{\rm d}t^2}= M_A U^{({\rm ext})}_A(\bold{z}_A)_{,i} 
+ \dfrac{1}{2} Q^{jk}_A U^{({\rm ext})}_A(\bold{z}_A)_{,ijk} + \O(\alpha^3),
\label{36-1987}
 \end{equation}
and hence we are ready to introduce in the external potential (cf. Eq. (\ref{16-1987}))
\begin{equation}
U^{({\rm ext})}_A(t,\bold{x})=\sum_{B \neq A} G \int_{V_B} \dfrac{\rho_B(t,\bold{X}_B)}
{\vert \bold{x}-\bold{z}_B-\bold{X}_B \vert} \,{\rm d}^3X_B, 
\label{37-1987}
\end{equation}
the second Taylor expansion we mentioned before, which is called multipole expansion, simply 
representing an expansion of $\vert \bold{x}-\bold{z}_B-\bold{X}_B \vert^{-1}$ about $\bold{X}_B=\bold{0}$, 
i.e., a power series in $\bold{X}_B$ having the form (like before 
$\dfrac{\partial}{\partial x^i} f\equiv f_{,i}$)
\begin{equation}
\dfrac{1}{\vert \bold{x}-\bold{z}_B-\bold{X}_B \vert}= \dfrac{1}{\vert \bold{x}-\bold{z}_B \vert} 
+ \left(\dfrac{1}{\vert \bold{x}-\bold{z}_B \vert}\right)_{,i} \left( - X^i_B \right) 
+ \dfrac{1}{2} \left(\dfrac{1}{\vert \bold{x}-\bold{z}_B \vert}\right)_{,ij} 
\left(-X^i_B\right) \left(-X^j_B\right) + \O(\vert \bold{X}_B\vert^3).
\label{39-1987}
\end{equation}
Thus, by substituting Eq. (\ref{39-1987}) in (\ref{37-1987}), the joint effect of the 
well-known result regarding the Laplacian
\begin{equation}
\triangle \left(\dfrac{1}{\vert \bold{x}-\bold{z}_B \vert}\right)	 
= - 4 \pi \delta^{(3)}\left( \bold{x}-\bold{z}_B\right),
\end{equation}
and of Eq. (\ref{Newtonian-Mass-Centred-Frame}) allows us to express
\begin{equation}
U^{({\rm ext})}_A(t,\bold{x}) = \sum_{B \neq A} G \left[ \dfrac{M_B}{\vert \bold{x}-\bold{z}_B \vert} 
+ \dfrac{1}{2}Q^{ij}_B  \left(\dfrac{1}{\vert \bold{x}-\bold{z}_B \vert}\right)_{,ij}  \right] + \O(\alpha^3),
\label{40-1987}
\end{equation}
$Q^{ij}_B$ denoting the quadrupole mass moment of the body $B$. The final step remaining consists in 
expressing (\ref{36-1987}) by means of (\ref{40-1987}). Since all derivatives of $U^{({\rm ext})}_A$ 
beyond the third order contribute $\O(\alpha^3)$ (see Eq. (\ref{alpha^3})), it is easy to see that 
the external motion admits the final $\alpha$-expanded form
\begin{equation}
\begin{split}
M_A \dfrac{{\rm d}^2 z_{i \; A}}{{\rm d}t^2} &= \sum_{B \neq A} \Biggl[GM_AM_B\dfrac{\partial}
{\partial z^i_A} \dfrac{1}{\vert \bold{z}_A-\bold{z}_B \vert}+ \dfrac{1}{2}G\left(M_A Q^{jk}_B
+ M_B Q^{jk}_A\right) \dfrac{\partial^3}{\partial z^i_A\partial z^j_A\partial z^k_A} 
\dfrac{1}{\vert \bold{z}_A-\bold{z}_B \vert} \Biggr] + \O(\alpha^3), \\
& \;\;\;\;\;\;\;\;\;\;\;\;\;\;\;\;\;\;\;\;\;\;\;\;\;\;\;\;\;\;\;\;\;\;\;\;\;\;\;\;\;\;\;\;\;
\;\;\;\;\;\;\;\;\;\;\;\;\;\;\;\;\;\;\;\;\;\;\;\;\;\;\;\;\;\;\;\;\;\;\;\;\;\;\;\;\;\;\;\;\;\;
\;\;\;\;\;\;\;\;\;\;\;\;\;\;\;\;\;\;\;\;\;\;\;\;\;\; (A=1,2,\dots,N).
\end{split}
\label{41-1987}
\end{equation}
The last equation clearly shows how the external Newtonian problem depends on the internal structure 
of bodies through the quadrupole moments. However, let us introduce the ellipticity parameter \cite{D1987,Fock1959}
\begin{equation}
\varepsilon \equiv\underset{A}{{\rm sup}} \left(\dfrac{\vert Q^{ij}_A\vert }
{\vert I^{ij}_A\vert} \right),
\end{equation}
which measures the relative deviation of the body $A$ from sphericity, in the sense that (cf. Eq. (\ref{33-1987}))
\begin{equation}
\varepsilon \sim \dfrac{\vert Q^{ij}_A \vert}{M_A L^2}.
\end{equation}
Under the hypothesis that the bodies, having finite mass and dimension, are almost spheric 
in shape (weakly self-gravitating bodies), i.e,
\begin{equation}
\varepsilon \ll 1,
\end{equation}
we have
\begin{equation}
Q^{ij} \sim \varepsilon M L^2 \ll 1.
\label{partial2}
\end{equation}
Furthermore, since the term
\begin{equation}
\left[\dfrac{\partial^3}{\partial z^i_A\partial z^j_A\partial z^k_A} 
\dfrac{1}{\vert \bold{z}_A-\bold{z}_B \vert}\right] \sim  \dfrac{1}{R^4}
\label{partial3}
\end{equation}
the joint effect of Eqs. (\ref{partial2}) and (\ref{partial3}) is such that the second term on the 
right-hand side of (\ref{41-1987}) gives a contribution
\begin{equation}
(G M) (\varepsilon M L^2)\left(\dfrac{1}{R^4}\right)= \left(\dfrac{GM^2}{R^2}\right) 
(\varepsilon \alpha^2) \ll \left(\dfrac{GM^2}{R^2}\right).
\end{equation}
In other words, the second term on the right-hand side of (\ref{41-1987}) is $\varepsilon \alpha^2$ 
smaller than the Newtonian force between two point masses, i.e., it is much smaller than the first 
term on the right-hand side of (\ref{41-1987}). Therefore, we can conclude that the internal structure 
is effaced in the description of the Newtonian external dynamics, which, upon discarding terms of 
order $\varepsilon \alpha^2$ and $\alpha^3$, turns out to be eventually represented by the autonomous 
system of $N$ differential equations
\begin{equation}
M_A \dfrac{{\rm d}^2 z_{i \; A}}{{\rm d}t^2} = \sum_{B \neq A} GM_AM_B\dfrac{\partial}
{\partial z^i_A} \dfrac{1}{\vert \bold{z}_A-\bold{z}_B \vert},
\;\;\;\;\;\;\;\;\;\;\;\;\;\;\;\;\ (A=1,2,\dots,N).
\end{equation}
describing the motion of the centre of mass $\bold{z}_A(t)$ of each body having constant mass $M_A$ \cite{D1987,Levi-Civita}. 

Regarding the internal motion (\ref{internal_problem}), it is possible to achieve an effacement of the 
external structure in a similar way as before by introducing a tidal expansion of the external potential 
(see Eq. (\ref{30-1987})) and  hence of the effective potential (\ref{6.3a}) according to
\begin{equation}
U^{{\rm (eff)}}_{A}(\bold{X}_A) = G^A_i \, X^i_A +\dfrac{1}{2}G^A_{ij}\, X^i_A X^j_A + \O(\vert \bold{X}_A\vert^3),
\end{equation} 
where the gravitational gradients or tidal moments \cite{D1} are defined by
\begin{equation}
 \begin{split}
G^A_i(t) & = U^{{\rm (ext)}}_{A}(\bold{z}_A)_{,i} - \dfrac{{\rm d}^2}{{\rm d}t^2}z_{i \; A}, \\
G^A_{i_1i_2\dots i_l}(t) & =U^{{\rm (ext)}}_{A}(\bold{z}_A)_{i_1i_2\dots i_l}, \;\;\;\;\;\; (l\geq2),
\label{Newtonian-Tidal-Moments}
\end{split}
\end{equation} 
and turn out to be automatically symmetric and trace free for the vanishing of the Laplacian of 
$U^{{\rm (ext)}}_{A}$ within the body $A$. In this case we find that the local coordinates (\ref{1.2a}) 
are such that the external potential gives a negligible contribution, in the sense that the effective potential 
is essentially reduced to tidal forces. Moreover, the internal problem is characterized by a particular 
issue (first noticed by Newton himself in his famous {\it Principia}) for which if the gravitational 
mass of a body equals its inertial mass then the external structure is even more effaced than one would 
expect \cite{D1987}. In fact, it has been this condition (implicitly always assumed) that has allowed us 
to define the effective potential (\ref{6.3a}). This peculiar property regarding the gravitational and 
inertial mass will become fundamental in general relativity and it is known as the weak equivalence principle.

\section{Relativistic tidal and multipole expansions} 

It should be clear from the Newtonian analysis outlined in the last section that, within the post-Newtonian 
framework, the following issues should be tackled:
\vskip 0.3cm
\noindent
(i) How to split the general $N$ body problem into two sub-problems concerning the internal and external dynamics ?
\vskip 0.3cm
\noindent
(ii) How can we achieve those effacement properties that would allow a decoupling of the two sub-problems?
\vskip 0.3cm
\noindent
(iii) Which is the analogue of the centre-of-mass-external-potential-effacing coordinates (\ref{1.2a})?
\vskip 0.3cm
\noindent
(iv) How can we generalize the Newtonian multipole and tidal moments (Eqs. (\ref{Newtonian-Multipole-moments}) 
and (\ref{Newtonian-Tidal-Moments}))?
\vskip 0.3cm
\noindent
Some of the above questions have been already answered in this paper and it should be also clear that they 
are all intertwined. In fact, the formalism developed by Damour, Soffel, and Xu in Refs. \citep{D1,D2,D3,D4} is 
such that the $N$ internal problems and the single external problem can be dealt with simultaneously within
the $N+1$ frames whose features have been illustrated before. Moreover, we have seen that the analogue of 
(\ref{1.2a}) is represented by (\ref{2.8aa}), which represents the starting point to achieve a 
{\it linear} (first order) description of the gravitational field (cf. Eqs. (\ref{3.11a}) and (\ref{4.3a})). 
We will shortly see how some of the components of the field $ e_{a}^{\; \mu} (X^0)$ occurring in (\ref{2.8aa}) 
are fixed though the search of some post-Newtonian effacement. Finally, the generalization of 
(\ref{Newtonian-Multipole-moments}) is represented by Blanchet-Damour mass and spin moments, whereas for Eq. 
(\ref{Newtonian-Tidal-Moments}) the role fulfilled by external potentials met in Eq. (\ref{4.7a}) will be crucial.

The first who dealt with the cancellation of internal structure at the first post-Newtonian order 
in the external problem was Levi-Civita \cite{Levi-Civita} (for a recent application of the
Levi-Civita model, motivated by the work in Refs. \cite{B1,B2,B3,B4}, 
see also Ref. \cite{B5}). Starting from the Einstein equations written in the 
de Donder(-Lanczos) gauge, Levi-Civita was able to split the Lagrangian for the geodesic motion of planets 
into a classical Newtonian part and into an Einstein modification. The required cancellation of the internal 
structure was then achieved by multiplying the Lagrangian by a constant giving rise to an equivalent set 
of equations of motion. Expressed in a modern language, 
the self-action effects were renormalised away and the final 
equations just ended up describing the dynamics of the (classically defined, through Eq. 
(\ref{Newtonian-Mass-Centred-Frame})) centre of mass of each body. This model has been surely overtaken 
by Damour, Soffel and Xu, who adopt a new purely post-Newtonian definition of centre-of-mass frame (see below) 
and, above all, leave always unspecified the form of the stress-energy tensor, in contrast with Levi-Civita 
approach, which uses a perfect matter model, although this choice is advocated only after a number of detailed 
calculations supplemented by some physical assumptions. However, in order to tackle the issue of generalizing 
the Newtonian description of Sec. III, we have to introduce Blanchet-Damour 
multipole moments. The great contribution of Ref. \cite{BD} and its generalization made in Refs. 
\citep{D1,D2,D3,D4} consists not only in having generalized the Newtonian moments seen in the last section, 
but also in having found a post-Newtonian analogue of both the (asymptotic) field multipole moments 
(\ref{40-1987}) (representing the gravitational field outside the material source and at a finite distance 
from it or close to space-like or null-like infinity, see Ref. \cite{Field}) and the source multipole 
moments (\ref{Newtonian-Mass-Moment}) and (\ref{Newtonian-Multipole-moments}) (expressing the 
moments in terms of the source). We have already seen that the gravitational field felt by each 
body both in the global frame and in its own frame can be decomposed as the sum of a locally generated 
part (defined in the harmonic gauge) and an external one. 
Blanchet-Damour multipole moments can be used to describe 
the former in {\it any} reference system. In other words, we can characterize either $w_\mu^{A}$ 
(i.e., the potential generated by one of the $N$ bodies of the system as seen in the global frame, 
cf. Eq. (\ref{3.29aa})) or $W_\alpha^{+\,A}$ (i.e., the local frame potential occurring in Eq. 
(\ref{4.7a})). Considering for definiteness the local potential $W_\alpha^{+\,A}=(W^{+\,A},W_a^{+\,A})$, 
it has been demonstrated that in any local $X^\alpha_A$ system it admits, outside the body $A$, 
the following Blanchet-Damour multipole expansion (label $A$ 
omitted on local time coordinate $T_A \equiv X^0_A/c$ 
and on spatial Minkowskian coordinates\footnote{Bearing in mind Eq. (\ref{3.6a}), spatial indices are 
always raised and lowered by means of Kronecker delta $\delta_{ab}$.} $X^a_A$) 
($\partial/\partial X^a_A \equiv \partial_a$) \cite{D1}:
\begin{equation}
\begin{split}
W^{+\,A}(T,X^a) & =G \Biggl\{  \left[ \dfrac{M_A(T+R/c)+M_A(T-R/c)}{2R}  \right] 
- \dfrac{\partial }{\partial X^{a_1}}\left[ \dfrac{M_A^{a_1}(T+R/c)
+M_A^{a_1}(T-R/c)}{2R}  \right] \\
& +\dfrac{1}{2!} \, \dfrac{\partial }{\partial X^{a_1}} \dfrac{\partial }{\partial X^{a_2}} 
\left[ \dfrac{M_A^{a_1a_2}(T+R/c)+M_A^{a_1a_2}(T-R/c)}{2R}  \right] \\
& + \dots + \dfrac{(-1)^l}{l!} \dfrac{\partial }{\partial X^{a_1}} \dfrac{\partial }
{\partial X^{a_2}} \dots \dfrac{\partial }{\partial X^{a_l}}\left[ \dfrac{M_A^{a_1a_2 \dots a_l}
(T+R/c)+M_A^{a_1a_2 \dots a_l}(T-R/c)}{2R}  \right] + \dots \Biggr\} \\
& + \dfrac{1}{c^2} \dfrac{\partial}{\partial T} \left( \Lambda^A - \lambda \right) + \O(c^{-4}),
\end{split}
\label{6.9aa}
\end{equation}
\begin{equation}
\begin{split}
W^{+\,A}_a(T,X^e) &= -G \Biggl\{-\dfrac{1}{R} \dfrac{{\rm d}}{{\rm d}t} M^A_a +\dfrac{1}{2!}\, 
\dfrac{\partial}{\partial X_{b_1}} \left( \dfrac{1}{R} \dfrac{{\rm d}}{{\rm d}t} M^A_{ab_1} \right)
-\dfrac{1}{3!} \, \dfrac{\partial}{\partial X_{b_1}}\dfrac{\partial}{\partial X_{b_2}} \left( 
\dfrac{1}{R} \dfrac{{\rm d}}{{\rm d}t} M^A_{a b_1b_2} \right)\\
& + \dots + \dfrac{(-1)^l}{l!} \dfrac{\partial}{\partial X_{b_1}}\dfrac{\partial}{\partial X_{b_2}} 
\dots \dfrac{\partial}{\partial X_{b_{l-1}}}  \left( \dfrac{1}{R} \dfrac{{\rm d}}{{\rm d}t} M^A_{a b_1b_2 
\dots b_{l-1}} \right) + \dots \\
& + \epsilon_{abc} \biggl[ -\dfrac{1}{2} \, \dfrac{\partial}{\partial X_b} 
\left(\dfrac{1}{R}S^c_A\right) + \dfrac{1}{2!}\dfrac{2}{3} \, \dfrac{\partial}{\partial X_b}
\dfrac{\partial}{\partial X^{d_1}} \left(\dfrac{1}{R}S^{c d_1}_A\right)-\dfrac{1}{3!}\dfrac{3}{4} \, 
\dfrac{\partial}{\partial X_b}\dfrac{\partial}{\partial X^{d_1}}\dfrac{\partial}
{\partial X^{d_2}} \left(\dfrac{1}{R}S^{c d_1d_2}_A\right) \\
& + \dots +\dfrac{(-1)^l}{l!} \dfrac{l}{l+1}  \dfrac{\partial}{\partial X_b}\dfrac
{\partial}{\partial X^{d_1}}\dfrac{\partial}{\partial X^{d_2}}\dots \dfrac{\partial}
{\partial X^{d_{l-1}}} \left(\dfrac{1}{R}S^{c d_1d_2 \dots d_{l-1}}_A\right) + \dots \biggr] \Biggr\} \\
& - \dfrac{1}{4}\dfrac{\partial}{\partial X^a} \left(\Lambda^A-\lambda \right) + \O(c^{-2}),
\end{split}
\label{6.9ba}
\end{equation}
where the sum over repeated indices is understood, $R=(X^a X_a)^{1/2}=(\delta_{ab}X^aX^b)^{1/2}$ 
and $\epsilon_{abc}$ denotes the Levi-Civita symbol and
\begin{equation}
\begin{split}
\Lambda^A & \equiv 4G \Biggl\{ \dfrac{1}{3}\,\left[ \dfrac{\mu^A(T+R/c)+\mu^A(T-R/c)}{2R} \right] 
-\dfrac{1}{2!}\dfrac{3}{5}\,\dfrac{\partial}{\partial X_{a_1}}\left[\dfrac{ 
\mu^A_{a_1}(T+R/c)+\mu^A_{a_1}(T-R/c)}{2R} \right] \\
& + \dots + \dfrac{(-1)^l}{(l+1)!}\dfrac{2l+1}{2l+3} \, \dfrac{\partial}{\partial X_{a_1}}
\dfrac{\partial}{\partial X_{a_2}} \dots \dfrac{\partial}{\partial X_{a_l}} 
\left[\dfrac{ \mu^A_{a_1a_2 \dots a_l}(T+R/c)+\mu^A_{a_1 a_2 \dots a_l}(T-R/c)}{2R} \right] + \dots \Biggr\},
\end{split}
\end{equation}
\begin{equation}
\mu^A_{a_1 a_2 \dots a_l} \equiv \int_{V_A} {\rm d}^3 X \,X^{\langle b} X^{a_1}X^{a_2} 
\dots X^{a_l \rangle} \Sigma_b (T,X^c).
\end{equation}
The functions $M^A_{a_1 a_2 \dots a_l}$ and $S^A_{a_1 a_2 \dots a_l }$ represent 
the Blanchet-Damour mass and spin multipole moments, respectively, 
which generalize the corresponding Newtonian 
quantities (\ref{Newtonian-Mass-Moment}) and (\ref{Newtonian-Multipole-moments}). 
They read in turn as \cite{BD,D1}
\begin{equation}
\begin{split}
M_A^{a_1 a_2 \dots a_l}(T) & \equiv \int_{V_A} {\rm d}^3X\, X^{\langle a_1}X^{a_2} 
\dots X^{a_l \rangle} \Sigma (T,X^c) + \dfrac{1}{2(2l+3)c^2}\dfrac{{\rm d}^2}{{\rm d}T^2}
\left(\int_{V_A} {\rm d}^3X  \, X^{\langle a_1}X^{a_2} \dots X^{a_l \rangle} 
X^bX_b \Sigma(T,X^c)\right) \\
& -\dfrac{4(2l+1)}{(l+1)(2l+3)c^2} \dfrac{{\rm d}}{{\rm d}T} \left( \int_{V_A} {\rm d}^3 X \,
X^{\langle b} X^{a_1}X^{a_2} \dots X^{a_l \rangle} \Sigma_b (T,X^c) \right), 
\;\;\;\;\;\; \;\;\; \;\;\;\;\;\; \;\;\;(l\geq0),
\end{split}
\label{6.11aa}
\end{equation}
\begin{equation}
\begin{split}
S_A^{c_1 c_2 \dots c_l}(T) & \equiv \int_{V_A} {\rm d}^3X \, \epsilon_{ab}^{\;\; \;
\langle c_l}X^{c_1}X^{c_2} \dots X^{c_{l-1} \rangle}X^a \Sigma^b (T,X^d),\;\;\;\;\;\; \;\;\; 
\;\;\;\;\;\; \;\;\; \;\;\;\;\;\; \;\;(l\geq1).
\end{split}
\label{6.11ba}
\end{equation}
These equations deserve some comments:
\vskip 0.3cm
\noindent
(i) The above relations are given in terms of compact-support integrals extended only 
over the volume $V_A$ of the {\it isolated} body $A$ and taken with respect to the origin of the local frame $A$.
\vskip 0.3cm
\noindent
(ii) In the sums occurring in Eqs. (\ref{6.9aa})--(\ref{6.11ba}) the spatial 
indices underlying the derivatives, the tensors and the local coordinates grow in number. 
In particular, in Eq. (\ref{6.9aa}) the sum begins with $l=0$, whereas (\ref{6.9ba}) with 
$l=1$. Moreover, all the spatial derivatives cannot be expressed with the concise multi-index 
notation, since they always represent derivatives of the {\it first} order with respect to 
the local coordinates $X^{a_i}$ ($i \in \{1,2,\dots,l\}$, $l \in \mathbb{N}-\{0\}$). However, 
we are aware of the fact that Eqs. (\ref{6.9aa})--(\ref{6.11ba}) become more succinct by 
employing Blanchet-Damour notation \citep{BD,D1,D2,D3,D4} (which represent a sort of generalized multi-index 
notation), but we have deliberately avoided using it in order to make all equations clearer 
(even if more lengthy) at a first sight.  
\vskip 0.3cm
\noindent
(iii) The first term occurring on the right-hand side of Eq. (\ref{6.11aa}) simply denotes the 
corresponding Newtonian quantities (\ref{Newtonian-Mass-Moment}) and (\ref{Newtonian-Multipole-moments}).
\vskip 0.3cm
\noindent
(iv) The term $l=0$ in Eq. (\ref{6.11aa}) is known as Blanchet-Damour mass of the body $A$.
\vskip 0.3cm
\noindent
(v) The spin multipole moments (\ref{6.11ba}) are purely Newtonian terms (i.e., they are written 
with Newtonian accuracy since no $\O(c^{-2})$ appear). In Ref. \cite{D3} a first-order post-Newtonian 
definition of the spin dipole $S^A_{c}$ (the term with $l=1$ in Eq, (\ref{6.11ba})) has been found. 
However, the post-Newtonian accuracy for all terms $l\geq 1$ of a closed self-gravitating system 
has been achieved in Ref. \cite{Damour-Iyer}.
\vskip 0.3cm
\noindent
(vi) $\lambda(T,X^a)$ denotes a gauge function taking into account the possibility of employing 
an arbitrary gauge. This aspect represents one of the generalizations introduced in Ref. \cite{D1} 
with respect to Ref. \cite{BD}, where the harmonic gauge was instead exploited. 
\vskip 0.3cm
\noindent
Having defined Blanchet-Damour mass and spin multipole moments, we can now generalize 
the remaining Newtonian notions, such as the definition of 
local centre of mass frame. In fact, a local $X^A_\alpha$ coordinate system will have its spatial origin 
coinciding for all $T_A$ times with the centre of mass of the body $A$ if its Blanchet-Damour dipole moment 
(i.e, the term $l=1$ of the sum (\ref{6.11aa}) vanishes. In other words, the post-Newtonian counterpart 
of Eq. (\ref{Newtonian-Mass-Centred-Frame}) now reads as 
\begin{equation}
\begin{split}
0=M_A^a(T_A) & = \int_{V_A} {\rm d}^3X_A \, X^a_A \, \Sigma(T_A,X^b_A) +\dfrac{1}{10c^2}
\dfrac{{\rm d}^2}{{\rm d}T_A^2}\int_{V_A} {\rm d}^3X_A \, X^a_A \, X^b_AX_b^A \, \Sigma(T_A,X^c_A) \\
& -\dfrac{6}{5c^2}\dfrac{{\rm d}}{{\rm d}T_A}\int_{V_A} {\rm d}^3X_A \,\left( X^a_A X_b^A  
-\dfrac{1}{3} \delta^a_{\,b} X^c_A X_c^A \right) \Sigma^b(T_A,X^e_A).
\end{split}
\label{5.10a}
\end{equation}
By virtue of the above equation, the abstract world line $\mathfrak{L}_A$ will now follow in a 
precise way the motion of the matter within $A$ and hence it can be identified as a centre-of-mass 
world line. However, how the definition (\ref{5.10a}) leads to the cancellation of the effects 
due to the external gravitational potential represents a more subtle argument than in Newtonian 
theory, where Eq. (\ref{1.2a}) makes the accelerated frames fulfil the role of both comoving 
frame and external field effacing frames (or freely falling frames) \cite{D1987}. In fact, if 
we regard (\ref{5.10a}) as identifying the comoving frames, we are left with some freedom in 
the choice of the system $X_\alpha^A$ which can be neatly exploited in order to efface the 
external gravitational potential also within the post-Newtonian framework \cite{D1}. 
Let us now explain how this goal can be achieved. 

By exploiting Eqs. (\ref{4.12da})--(\ref{4.55a}), the external potential occurring in (\ref{4.7a}) 
can be completely decomposed in terms of the local mass density $\Sigma^B_{\beta}(X_B)$ as \cite{D1}
\begin{equation}
\overline{W}^{A}_{\alpha}(X_A) = \sum_{B \neq A} W^{B/A}_\alpha(x^\mu(X_A))
+\overline{W}^{\prime \prime \, A}_\alpha(X_A),
\label{5.8ca}
\end{equation}
where the first term represents the contribution resulting from all bodies $B \neq A$
\begin{equation}
W^{B/A}_\alpha(x^\mu(X_A))=\square^{-1}_{x,{\rm sym}} \left( -4 \pi G \left\vert 
\dfrac{\partial X_B(x)}{\partial x}\right\vert \mathcal{A}^{A\,(-1)}_{\,\alpha \mu}(T_B) \, 
\mathcal{A}^{B}_{\, \mu \beta}(T_B) \,\Sigma^B_{\beta}(X_B(x)) \right),
\label{5.8da}
\end{equation}
whereas $\overline{W}^{\prime \prime \, A}_\alpha(X_A)$ results from the inertial effects 
underlying the change of (``accelerated'') frames $x^\mu \rightarrow X^\alpha_A$ and reads as
\begin{equation}
\overline{W}^{\prime \prime \, A}_\alpha(X_A)= - \mathcal{A}^{A\,(-1)}_{\,\alpha \mu}(T_A) 
\mathcal{B}^A_\mu(X_A).
\label{5.8ea}
\end{equation}
Note that in the above equations the term $\mathcal{A}^{A\,(-1)}$ results from the inversion 
of Eq. (\ref{4.12da}) (i.e., Eq. (\ref{4.13aa})), while the Jacobian occurring in (\ref{5.8da}) 
relates the global density of $B$ to its local frame through
\begin{equation}
\sigma^B_\mu(x)=\left \vert \dfrac{\partial X_B(x)}{\partial x}\right\vert  
\mathcal{A}^{B}_{\, \mu \beta}(X^0_B) \,\Sigma^B_{\beta}(X_B(x)).
\end{equation}  
Using the definition (\ref{5.8ca}), we can say that the external potential is locally effaced in 
the frame of the body $A$ if it vanishes for all times $T_A$ at the origin of the frame \cite{D1}, i.e.,
\begin{equation}
\overline{W}^{A}_{\alpha}(T_A,0,0,0)=0, \;\;\;\;\;\;\;\;\;\;\;\; \forall\,T_A.
\label{5.12a}
\end{equation}
Because of the inertial terms appearing in Eq. (\ref{5.8ea}), the four effacement conditions 
(\ref{5.12a}) involve a choice of some of the $e_{a}^{\; \mu} (X^0)$ coefficients occurring in 
Eq. (\ref{2.8aa}), as we pointed out before (see Ref. \cite{D1} for more details). 

In order to conclude the ``generalization task'' of this section, let us generalize to the 
post-Newtonian framework the tidal moments (\ref{Newtonian-Tidal-Moments}). Let in some local 
frame ($\partial_a=\partial/\partial X^a_A$, $\partial_T=c\,\partial/\partial X^0_A$)
\begin{equation}
\begin{split}
& {\overline{E}}^A_a(T,X^b) \equiv \partial_a \overline{W}^{A} +\dfrac{4}{c^2} 
\partial_T \overline{W}^{A}_a, \\
& {\overline{B}}^A_a(T,X^d) \equiv \epsilon_{abc} \partial_b \left(-4 \overline{W}^A_c \right),
\end{split}
\end{equation}
be the external electric and magnetic gauge invariant fields. The gravitoelectric and gravitomagnetic 
post-Newtonian tidal moments are defined as \citep{D1,D2,D3,D4}
\begin{equation}
\begin{split}
& G^A(T) \equiv \overline{W}^A(T,0,0,0) ,\\
& G^A_{a_1a_2\dots a_l}(T)\equiv  \left. \partial_{\langle a_1} \partial_{a_2} \dots \partial_{a_{l-1}} 
\overline{E}^A_{a_l \rangle} (T,\bold{X}) \right \vert_{X^b=0}, \;\;\;\;\;\; \;\;\;\;\; (l \geq 1), \\
& H^A_{a_1a_2\dots a_l}(T) \equiv \left. \partial_{\langle a_1} \partial_{a_2} \dots \partial_{a_{l-1}} 
\overline{B}^A_{a_l \rangle} (T,\bold{X}) \right \vert_{X^b=0}, \;\;\;\;\;\;\;\;\;\;\;  (l \geq 1).
\end{split}
\label{6.13a}
\end{equation}
respectively. Note however that the monopole tidal moment $G^A(T)$ can be gauged away through 
Eq. (\ref{5.12a}). Eventually, by means of Eqs. (\ref{6.9aa}), (\ref{6.9ba}), and (\ref{5.8ca}) 
it is possible to express the post-Newtonian tidal moments of the body $A$ through the superposition of 
$N$ contributions: $N-1$ of them are generated separately by each body $B \neq A$ of the system while 
the last one represents the inertial terms. All of them are completely computable from Blanchet-Damour 
multipole moments \cite{D2}.
 
\section{The monopole model} 
 
The starting point towards the derivation of the simplest Lagrangian model describing the dynamics of the 
$N$-body system is represented by Blanchet-Damour multipole moments (\ref{6.11aa}) and (\ref{6.11ba}) and the 
tidal moments (\ref{6.13a}). Bearing in mind that the vanishing of the four-divergence of the 
energy-momentum tensor $T^{\alpha \beta}(X^\gamma_A)=0$ in the local $X^\alpha_A$ frame implies, in the context of linearized gravity, the conservation for all times $T_A$ of the mass monopole, 
mass dipole and spin dipole moments of an isolated system, the post-Newtonian counterpart of this 
result leads to the existence of constraints on the time evolution of the three lowest local Blanchet-Damour 
multipoles of the body $A$ having the form \citep{D1,D2,D3,D4}
\begin{equation}
\dfrac{{\rm d}}{{\rm d}T_A}M^A=\dfrac{1}{c^2}F_{0}^{({\rm 1PN})}\left(M^{(p)A}_{a_1 a_2 
\dots a_l},G^{(p^{\prime})A}_{b_1 b_2 \dots b_l}\right) + \O(c^{-4}),
\label{7.1aa}
\end{equation}  
\begin{equation}
\begin{split}
\dfrac{{\rm d}^2}{{\rm d}T^2_A}M^A_a & =M_A\,G^A_a+M_A^{b_1}G^A_{ab_1}+\dfrac{1}{2!}M_A^{b_1 b_2}
G^A_{ab_1b_2}+ \dots + \dfrac{1}{l!}M_A^{b_1 b_2\dots b_l}G^A_{ab_1b_2\dots b_l}+ \dots \\
& +\dfrac{1}{c^2} F_{a}^{({\rm 1PN})}\left(M^{(p)A}_{a_1 a_2 \dots a_l},S^{(q)A}_{b_1 b_2 \dots b_l};
G^{(p^{\prime})A}_{a^{\prime}_1 a^{\prime}_2 \dots a^{\prime}_l},H^{(q^{\prime})A}_{b^{\prime}_1 
b^{\prime}_2 \dots b^{\prime}_l}\right) +\O(c^{-4}),
\end{split}
\end{equation}
\begin{equation}
\dfrac{{\rm d}^2}{{\rm d}T^2_A}S^A_a= \epsilon_{a}^{\;\,bc} \left(M^A_b G^A_c+ M^A_{bd_1} G^{Ad_1}_{c}
+\dfrac{1}{2!}M^A_{bd_1d_2} G^{Ad_1d_2}_{c}+\dots+ \dfrac{1}{l!}M^A_{bd_1d_2\dots d_l} G^{Ad_1d_2
\dots d_l}_{c}+ \dots \right) + \O(c^{-2}),
\label{7.1ca}
\end{equation}
with
\begin{equation}
M^{(p)A}_{a_1 a_2 \dots a_l}=\dfrac{{\rm d}^p}{{\rm d}T^p_A} M^A_{a_1 a_2 \dots a_l}, \;\;\;\;\; {\rm etc.}
\end{equation}
$F_{0}^{({\rm 1PN})}$ and $F_{a}^{({\rm 1PN})}$ are referred to as energy loss and post-Newtonian 
force terms, respectively. The right-hand sides of Eqs. (\ref{7.1aa})--(\ref{7.1ca}) are all given 
in terms of bilinear couplings between the Blanchet-Damour mass and spin multipole moments of the body $A$ and the 
post-Newtonian tidal moments felt by $A$ and their time derivatives. In particular, they are 
separately linear, on the one hand, on $M^{A}_{a_1 a_2 \dots a_l}$, $S^{A}_{a_1 a_2 \dots a_l}$, 
and their time derivatives and, on the other hand, on $G^{A}_{a_1 a_2 \dots a_l}$, 
$H^{A}_{a_1 a_2 \dots a_l}$, and their time derivatives.

The most feasible framework that can be built up within Damour, Soffel, and Xu formalism is 
represented by the monopole model, according to which each body is described exclusively by its 
Blanchet-Damour mass and the whole system is subjected to the ansatz 
\begin{equation}
\begin{split}
& M^A_{a_1}=M^A_{a_1 a_2}=\dots=M^A_{a_1 a_2 \dots a_l}=\dots=0,\\
& S^A_{a_1}=S^A_{a_1 a_2}=\dots=S^A_{a_1 a_2 \dots a_l}=\dots=0,\\
\end{split}
\end{equation}
which in turn, jointly with the arguments formulated at the end of Sec. IV  
and the bilinear structure of Eq. (\ref{7.1aa})--(\ref{7.1ca}), implies
\begin{equation}
\dfrac{{\rm d}}{{\rm d}T_A}M^A=\O(c^{-4}).
\end{equation}
In other words, the monopole model is defined to have constant Blanchet-Damour 
mass while all other Blanchet-Damour moments vanish 
(and hence the post-Newtonian tidal moment, too). Within this framework, the 
global-frame post-Newtonian equations of motion for a system of $N$ (weakly 
self-gravitating) Blanchet-Damour monopoles leads to 
the well-known Einstein-Infeld-Hoffmann model, in the sense that the acceleration 
of each body is given \cite{D1,D4,E1,E2}
\begin{equation}
\dfrac{{\rm d}^2}{{\rm d}t^2}z^i_A =a^{i({\rm EIH})}_A(\bold{z}_B,\bold{v}_B)+\O(c^{-4}),
\end{equation}
with
\begin{equation}
\begin{split}
\bold{a}_A^{({\rm EIH})} &= - \sum_{B \neq A} \dfrac{G M_B}{r^2_{AB}} \bold{n}_{AB} 
-\dfrac{1}{c^2} \sum_{B \neq A} \dfrac{G M_B}{r^2_{AB}} \bold{n}_{AB} \Biggl[\bold{v}^2_A 
+ 2 \bold{v}^2_B - 4 \bold{v}_A \cdot \bold{v}_B -\dfrac{3}{2} (\bold{n}_{AB} \cdot \bold{v}_B)^2 
-4 \sum_{C \neq A} \dfrac{GM_C}{r_{AC}} \\
& -   \sum_{C \neq B} \dfrac{GM_C}{r_{BC}} \left(  1 + \dfrac{1}{2} \dfrac{r_{AB}}{r_{CB}}\, 
\bold{n}_{AB} \cdot \bold{n}_{CB} \right) \Biggr]-\dfrac{1}{c^2}\,\dfrac{7}{2} 
\sum_{B \neq A} \sum_{C \neq B} \bold{n}_{BC} \dfrac{G^2 M_B M_C}{r_{AB}r^2_{BC}} \\
& + \dfrac{1}{c^2} \sum_{B \neq A} (\bold{v}_A-\bold{v}_B) \dfrac{GM_B}{r^2_{AB}}
(4 \bold{n}_{AB}\cdot \bold{v}_A - 3 \bold{n}_{AB} \cdot \bold{v}_B),
\end{split}
\end{equation}
where
\begin{equation}
\begin{split}
& r_{AB} = \vert \bold{z}_A(t)-\bold{z}_B(t)\vert,\\
& \bold{n}_{AB} = \dfrac{\vert \bold{z}_A(t)-\bold{z}_B(t)\vert}{r_{AB}}.
\end{split}
\end{equation}
 
\section{Analysis of the integro-differential dynamical equations}
 
The formalism outlined in Secs. II-V has the great advantage of providing 
an efficient analysis where the integro-differential nature of the dynamical equations underlying 
the description of the $N$-body system can be ``tackled'' 
through the introduction of the Blanchet-Damour 
multipole moments and the post-Newtonian tidal moments. In this Section we aim at outlining 
the features of the post-Newtonian equations describing the system without introducing any expansion,
the reason being that a powerful computational recipe is not a substitute for the matters of
principle that the theoretical physicist must consider in his research.

The equations expressing the evolution of the material distribution of each body can be achieved 
in the following way \cite{D1}. At the first post-Newtonian order the equations representing the 
exchange of energy and momentum between each volume element of the system and the gravitational field, i.e., 
\begin{equation}
T^{\mu \nu}_{\;\;\;\; ; \nu}=0,
\end{equation}
(the semicolon indicating the usual covariant derivative) are found to be equivalent, in the local $A$-frame, to the evolution system for 
$\Sigma^\alpha=(\Sigma,\Sigma^a)$ (cf. Eqs. (\ref{4.14a})  and (\ref{4.4a})) (label $A$ omitted)
\begin{equation}
\renewcommand{\arraystretch}{2.0}
\begin{dcases}
& \dfrac{\partial}{\partial T}\left[ \left(1 + \dfrac{4}{c^2}W \right)\Sigma^a\right] 
+\dfrac{\partial}{\partial X^b} \left[\left(1 + \dfrac{4}{c^2}W \right)T^{ab}\right] 
= F^a(T,\bold{X}) + \O(c^{-4}), \\
& \dfrac{\partial}{\partial T} \Sigma + \dfrac{\partial}{\partial X^a} \Sigma^a 
= \dfrac{1}{c^2} \dfrac{\partial}{\partial T} T^{bb}-\dfrac{1}{c^2} 
\Sigma \dfrac{\partial}{\partial T} W +  \O(c^{-4}),
\end{dcases} \\ [2em]
\label{5.6a}
\end{equation}
where $F^a(T,\bold{X}) $ resembles the Lorentz force density of electromagnetism
\begin{equation}
F^a = \Sigma E^a (W) +\dfrac{1}{c^2} B^{a}_{\,b}(W) \Sigma^b,
\end{equation}
fulfilling the role of a gravitational force density at the first post-Newtonian order, as 
is clear from Eq. (\ref{5.6a}). The set of Eqs. (\ref{5.6a}) should be supplemented with 
the equations expressing the evolution of the gauge invariant gravitational potential 
$W^\alpha$, Eqs. (\ref{4.5a}) and (\ref{5.8ca})--(\ref{5.8ea}), which make the whole system 
nonlocal in space because of the inverse operator $\square^{-1}$ (which is an integral
operator with kernel given by a Green function), thus recovering its 
integro-differential nature that we mentioned before.

First of all, we rewrite Eqs. (\ref{5.6a}) as
\begin{equation}
\begin{split}
W \dfrac{\partial \Sigma^a}{\partial T}&= - \Sigma^a \dfrac{\partial W}{\partial T} 
-\dfrac{c^2}{4} \dfrac{\partial \Sigma^a}{\partial T} - \dfrac{c^2}{4}\dfrac{\partial}{\partial X^b} 
\left[\left(1 + \dfrac{4}{c^2}W \right)T^{ab}\right] +\dfrac{c^2}{4} F^a + \O(c^{-4}), \\
\dfrac{\partial W}{\partial T}  &= \dfrac{1}{\Sigma} \dfrac{\partial T^{bb}}{\partial T} 
- \dfrac{c^2}{\Sigma} \dfrac{\partial \Sigma}{\partial T}-\dfrac{c^2}{\Sigma} 
\dfrac{\partial \Sigma^a}{\partial X^a} + \O(c^{-4}),
\end{split}
\end{equation}
which, upon defining
\begin{equation}
\begin{split}
f(T,\bold{X})& \equiv\dfrac{1}{\Sigma} \dfrac{\partial T^{bb}}{\partial T} - \dfrac{c^2}{\Sigma} 
\dfrac{\partial \Sigma}{\partial T}-\dfrac{c^2}{\Sigma} \dfrac{\partial \Sigma^a}
{\partial X^a} + \O(c^{-4}), \\
h^a(T,\bold{X})& \equiv -\dfrac{c^2}{4} \dfrac{\partial \Sigma^a}{\partial T} 
- \dfrac{c^2}{4}\dfrac{\partial}{\partial X^b} \left[\left(1 + \dfrac{4}{c^2}W \right)
T^{ab}\right] +\dfrac{c^2}{4} F^a + \O(c^{-4}),
\end{split}
\end{equation}
assumes the concise equivalent form
\begin{equation}
\renewcommand{\arraystretch}{2.0}
\begin{dcases}
& W \dfrac{\partial \Sigma^a}{\partial T}  = - \Sigma^a f + h^a, \\
& \dfrac{\partial W}{\partial T}  = f.
\end{dcases} \\ [2em]
\label{5.6a_concise}
\end{equation}
The distributional relation involving the step function and Dirac delta function
\begin{equation}
\dfrac{\partial}{\partial T} \theta(T) = \delta (T),
\end{equation}
is such that Eq. (\ref{5.6a_concise}b) can be solved at once by writing
\begin{equation}
W(T,\bold{X})= \int {\rm d} \tau \, \theta(T-\tau) f (\tau,\bold{X}).
\label{mia1}
\end{equation}

At this stage, we can describe concisely the methods that might lead to the  
resolution of the integro-differential system (\ref{5.6a_concise}). First of all, bearing in 
mind Eqs. (\ref{4.7a}), (\ref{4.5a}), and (\ref{mia1}) we can easily derive that 
\begin{equation}
\square_X W(T,\bold{X}) = \int {\rm d}\tau \,  \square_X  \left[\theta(T-\tau)
f (\tau,\bold{X}) \right]= \square_X \left(  W^{+} + \overline{W} \right) 
= - 4 \pi G \Sigma(T,\bold{X}),
\label{mia2}
\end{equation}
since $\square_X  \overline{W}=0$. Therefore, the above equation can give some 
information on the features of the function $f (\tau,\bold{X})$ and can also be worked out 
by exploiting the following relation involving the d'Alembert operator of two 
{\it scalar} functions $g=g(T,\bold{X})$ and $z=z(T,\bold{X})$:
\begin{equation}
\square_X (gz) = \left(\square_X g\right)z+g\left(\square_X z\right)
+2 \eta_{\alpha \beta} \partial^{\alpha} g \, \partial^{\beta} z.
\label{Box_property}
\end{equation}
Moreover, another way to proceed can be as follows. Equations (\ref{4.7a}) and (\ref{4.5a}) 
suggest applying the d'Alembert operator to the system (\ref{5.6a_concise}) and exploit, 
whenever we meet $\square_X W(T,\bold{X})$, the fact that its locally generated part $W^{+}_{\alpha}$ is such that
\begin{equation}
\square_X W^{+}_{\alpha} = - 4 \pi G \Sigma_{\alpha}.
\end{equation}
Unfortunately, this approach exhibits many drawbacks due mainly to the fact that the nice 
property (\ref{Box_property}) cannot always be exploited since in Eq. (\ref{5.6a_concise}) 
both vector and tensor quantities appear. As an example, 
for a vector $\bold{A}$ the vector Laplacian is defined as
\begin{equation}
\triangle_X \bold{A}= \bold{\nabla}_X \left( \bold{\nabla}_X \cdot \bold{A} \right) 
- \bold{\nabla}_X \times \left( \bold{\nabla}_X \times \bold{A} \right),
\end{equation}
which reduces to the usual scalar Laplacian when applied to scalar fields. Since the vector Laplacian 
returns a vector field (unlike the scalar Laplacian, which gives a scalar quantity), 
Eq. (\ref{Box_property}) cannot be exploited throughout the calculation we delineated above. 
In spite of this, we can easily evaluate $\square_X  f (T,\bold{X})$, which may represent a 
crucial step toward the resolution of (\ref{5.6a_concise}), since it leads to an useful 
relation to be employed in Eq. (\ref{mia2}). First of all, we can write $f(T,\bold{X})$ 
in an equivalent form as (cf. Eq. (\ref{4.4a}))
\begin{equation}
f(T,\bold{X})= \Sigma^{-1} \left( \dfrac{\partial}{\partial T} T^{00} 
-c^2 \bold{\nabla}_X \cdot \bold{\Sigma} \right),
\label{f_function}
\end{equation}
which does not represent the product of two scalar functions because of the presence of 
the component $T^{00}$ of the stress-energy tensor $T^{\mu \nu}$. However we can easily 
show that, since in Minkowski background the d'Alembert operator, when acting on the first term 
in round brackets of (\ref{f_function}), behaves as if it was applied on a scalar function, 
Eq. (\ref{Box_property}) can be still exploited. Indeed, in the most general curved background 
geometry we should have evaluated the following quantity\footnote{We are performing a general 
calculation and hence Greek indices are written without following the distinction between 
local and global charts outlined at the beginning of Sec. II.}
\begin{equation}
\begin{split}
\square \left( \partial_\lambda T^{\mu \nu}\right) & = g^{\alpha \beta} \left( \partial_\lambda 
T^{\mu \nu}\right)_{;  \beta \alpha } = g^{\alpha \beta} \left(T^{\mu \nu}_{\;\;\;\; ;\lambda} 
- \Gamma^{\mu}_{\; \sigma \lambda} T^{\sigma \nu}-\Gamma^{\nu}_{\; \sigma \lambda} 
T^{ \mu \sigma} \right)_{;  \beta \alpha } \\
&= g^{\alpha \beta} \left[ T^{\mu \nu}_{\;\;\;\; ;\lambda \beta \alpha} 
- \left( \Gamma^{\mu}_{\; \sigma \lambda} T^{\sigma \nu} \right)_{; \beta \alpha} 
- \left( \Gamma^{\nu}_{\; \sigma \lambda} T^{ \mu \sigma} \right)_{;  \beta \alpha } \right].
\end{split}
\label{General_Box}
\end{equation}
In order to evaluate the above equation, 
we only need to consider that a third-order covariant derivative of a twice contravariant tensor reads as
\begin{equation}
T^{\mu \nu}_{\;\;\;\; ;\lambda \beta \alpha} = \partial_\alpha \left( 
T^{\mu \nu}_{\;\;\;\; ;\lambda \beta}\right) - \Gamma^{\sigma}_{\; \alpha \beta} 
T^{\mu \nu}_{\;\;\;\; ;\lambda \sigma} -\Gamma^{\sigma}_{\; \alpha \lambda} 
T^{\mu \nu}_{\;\;\;\; ; \sigma \beta}+ \Gamma^{\mu}_{\; \sigma \alpha } 
T^{\sigma \nu}_{\;\;\;\; ;\lambda \beta} + \Gamma^{\nu}_{\; \sigma \alpha } 
T^{\mu \sigma}_{\;\;\;\; ;\lambda \beta},   
\end{equation}
whereas a second order one is given by
\begin{equation}
\begin{split}
T^{\mu \nu}_{\;\;\;\; ;\lambda \beta} &= \partial_\beta \left( T^{\mu \nu}_{\;\;\;\; ;\lambda}\right) 
- \Gamma^{\epsilon}_{\; \lambda \beta } T^{\mu \nu}_{\;\;\;\; ;\epsilon} 
+ \Gamma^{\mu}_{\; \epsilon \beta } T^{\epsilon \nu}_{\;\;\;\; ;\lambda} 
+ \Gamma^{\nu}_{\; \epsilon \beta } T^{\mu \epsilon}_{\;\;\;\; ;\lambda} \\
& = \partial_\beta \left( \partial_\lambda T^{\mu \nu} + \Gamma^{\mu}_{\; \xi \lambda } T^{\xi \nu} 
+ \Gamma^{\nu}_{\; \xi \lambda } T^{\mu \xi} \right) -\Gamma^{\epsilon}_{\; \lambda \beta } 
\left(  \partial_\epsilon T^{\mu \nu} + \Gamma^{\mu}_{\; \epsilon \xi  } T^{\xi \nu} 
+ \Gamma^{\nu}_{\; \epsilon \xi} T^{\mu \xi} \right)\\
& + \Gamma^{\mu}_{\; \epsilon \beta } \left(\partial_\lambda T^{\epsilon \nu}
+ \Gamma^{\epsilon}_{\; \lambda \xi  } T^{\xi \nu} + \Gamma^{\nu}_{\; \lambda \xi} 
T^{\epsilon \xi} \right) + \Gamma^{\nu}_{\; \epsilon \beta } \left(\partial_\lambda 
T^{\mu \epsilon }+ \Gamma^{\mu}_{\; \xi \lambda   } T^{\xi \epsilon} 
+ \Gamma^{\epsilon}_{\; \xi \lambda} T^{\mu \xi} \right).
\end{split} 
\end{equation}
At this stage, bearing in mind the last two equations, the simple argument according to which the 
connection coefficients along with their partial derivatives constitute functions which can always 
be made to vanish in flat Minkowski space (e.g. by employing the global rectangular patch 
$(t,x,y,z)$) allows us to conclude that, in the flat background geometry we are handling, 
Eq. (\ref{General_Box}) reduces simply to
\begin{equation}
\square \left( \partial_\lambda T^{\mu \nu}\right) = \eta^{\alpha \beta}
\partial_\alpha \partial_\beta \partial_\lambda T^{\mu \nu},
\end{equation}
actually showing that the action of $\square_X$ on $\dfrac{\partial}{\partial T}  T^{00}$ 
is the same as on generic scalar functions.

Therefore, we can now employ (\ref{Box_property}) and write
\begin{equation}
\begin{split}
\square_X f(T,\bold{X}) & = \left( \square_X \Sigma^{-1}\right)  \left( \dfrac{\partial}
{\partial T} T^{00} -c^2 \bold{\nabla}_X \cdot \bold{\Sigma} \right)+ \Sigma^{-1} \left[   
\dfrac{\partial}{\partial T} \left( \square_X T^{00}\right) -c^2 \square_X 
\left(\bold{\nabla}_X \cdot \bold{\Sigma} \right) \right] \\
& +2 \eta_{\alpha \beta} \partial^{\alpha }\left(\Sigma^{-1}\right)\partial^{\beta} 
\left( \dfrac{\partial}{\partial T} T^{00} -c^2 \bold{\nabla}_X \cdot \bold{\Sigma} \right),
\end{split}
\end{equation}
where we have employed the commutator relation $\left[\dfrac{\partial}{\partial T},\square_X\right]=0$.

Regrettably, it is clear that both approaches outlined above, despite being ultimately 
intertwined, are really hard to pursue. 

\section{Beyond two bodies: the restricted three-body problem}

Following the review spirit of this paper, we have decided to conclude it with a section 
dealing with the analysis of the relativistic restricted three-body problem which, 
as shown in Ref. \cite{Brumberg2003}, in the simplest case of (quasi-)circular motion and 
within the $c^{-5}$ order shows important deviations from the Newtonian theory because of 
the loss of energy by gravitational radiation. In particular, Lagrangian points become 
Lagrange-like quasi-libration points undergoing secular trends. This approach differs a little 
bit from the one described in the previous sections, as we will shortly see.

Consider a framework for which the gravitational field is described in terms of a global chart by the metric
\begin{equation}
g_{\mu \nu}= \eta_{\mu \nu}+ h_{\mu \nu},
\label{Brumberg2.2}
\end{equation}
whereas the motion of a test particle in such a field is given by the geodesic equation
\begin{equation}
\ddot{x}^{\alpha} + \Gamma^{\alpha}_{\; \mu \nu} \dot{x}^{\mu} \dot{x}^{\nu}=0,
\label{Brumberg2.3}
\end{equation}
the dot denoting differentiation with respect to the coordinate time $t$. In order to achieve 
a $c^{-5}$ accuracy, we should evaluate Christoffel symbols appearing in (\ref{Brumberg2.3}) within the accuracy 
\begin{equation}
\Gamma^{i}_{\; 00} \sim c^{-7}, \;\;\;\;\Gamma^{i}_{\; 0k}, \Gamma^{0}_{\; 00} \sim 
c^{-6},\;\;\;\;\Gamma^{i}_{\; km}, \Gamma^{0}_{\; 0k} \sim c^{-5}, \;\;\;\; \Gamma^{0}_{\; km} \sim c^{-4},
\end{equation}
and introduce the usual post-Newtonian approximation  
\begin{equation}
\begin{split}
h_{00} &= c^{-2} h^{(2)}_{00}+c^{-4} h^{(4)}_{00}+c^{-5} A^{(5)}_{00}
+c^{-6} h^{(6)}_{00}+c^{-7} A^{(7)}_{00}+ \O(c^{-8}),\\
h_{0i} &= c^{-3} h^{(3)}_{0i}+c^{-5} h^{(5)}_{0i}+c^{-6} A^{(6)}_{0i}+ \O(c^{-7}),\\
h_{ij} &= c^{-2} h^{(2)}_{ij}+c^{-4} h^{(4)}_{ij}+c^{-5} A^{(5)}_{ij}+ \O(c^{-6}),
\end{split}
\end{equation}
the components $A^{(5)}_{00}$, $A^{(7)}_{00}$, $A^{(6)}_{0i}$, and $A^{(5)}_{ij}$ being responsible 
of the gravitational radiation of celestial bodies (i.e., second-half post-Newtonian terms). 
By employing the above expansions and the harmonic gauge, it is possible to show that the Lagrangian 
$L$ describing the dynamics of a test particle in the gravitational field (\ref{Brumberg2.2}) 
within the order $\O(c^{-5})$ is given by the function
\begin{equation}
L=f(x^{\mu},h^{(n)}_{\mu \nu},A^{(m)}_{\mu \nu}),
\label{Brumberg3.5}
\end{equation}
whose particular form can be read in Ref. \cite{Brumberg2003}. 

In order to describe the quasi-Newtonian restricted quasi-circular three-body problem, we should 
first of all consider the dynamics of the two primaries generating the field in which the motion 
of the small planetoid takes place. By employing the barycentric system (defined classically by 
(\ref{Newtonian-Mass-Centred-Frame})) the dynamical equations, to first order, of the two bodies 
of masses $M_1$ and $M_2$ are given by \cite{Brumberg2003}
\begin{equation}
M_1 x_1^i+M_2 x_2^i+\dfrac{1}{2c^2} \left[M_1 \left((\bold{v}_1)^2-\dfrac{GM_2}{R}\right)x_1^i
+M_2 \left((\bold{v}_2)^2-\dfrac{GM_1}{R}\right)x_2^i \right]+ \O(c^{-4})=0,
\end{equation}
and 
\begin{equation}
M_1 v_1^i+M_2 v_2^i+\dfrac{1}{2c^2} \left[M_1 \left((\bold{v}_1)^2-\dfrac{GM_2}{R}\right)v_1^i
+M_2 \left((\bold{v}_2)^2-\dfrac{GM_1}{R}\right)v_2^i-\dfrac{G M_1 M_2}{R}\left(\bold{N}\cdot\bold{v}_1 
+ \bold{N}\cdot\bold{v}_2\right)N^i \right]+ \O(c^{-4})=0,
\end{equation}
where
\begin{equation}
\begin{split}
& R^i \equiv x_1^i-x_2^i, \;\;\;\; V^i \equiv v_1^i-v_2^i, \;\;\;\; v^i_K \equiv \dot{x}^i_K,\\
& R \equiv \sqrt{R_i R^i}, \;\;\;\; N^i \equiv \dfrac{R^i}{R},\;\;\;\; \bold{N}\cdot\bold{v}=\delta_{ij}N^i v^j,
\end{split}
\end{equation}
(the index $K=1,2$ labelling the primaries). In terms of relative coordinates, the equations of 
motion can be put in the form \cite{Brumberg2003}
\begin{equation}
\ddot{R}_i= B_i \equiv B^{(0)}_i+c^{-2}B^{(2)}_i +c^{-4}B^{(4)}_i +c^{-5}B^{(5)}_i +\O({c^{-6}}),
\label{Brumberg4.10}
\end{equation}
where the coefficients $B^{(n)}_i$ are functions of the relative coordinates $R^i$, $V^i$ and the 
masses $M_1$ and $M_2$ of the bodies. In particular, the term $B^{(5)}_i$ is a dissipative term having the form
\begin{equation}
B^{(5)}_i=\dfrac{8}{5}\dfrac{G^2 M^2 \mu}{R^3} \left[ \left( 3 \bold{V}^2 +\dfrac{17}{3}
\dfrac{GM}{R}\right) \left(N^kV_k\right)N^i -\left(\bold{V}^2 +\dfrac{3GM}{R}\right)V^i\right],
\end{equation}
$\mu \equiv M_1M_2/(M_1+M_2)^2$ being a dimensionless parameter. Such a term is responsible for the 
secular changes in the semi-major axes and eccentricities of the orbits of the primaries, 
the resulting secular variation in the orbital periods being a well-confirmed effect 
in the context of binary pulsars. 

In order to deal with the restricted three-body problem, we take into account the simplest situation 
in which the two massive bodies move along quasi-circular orbits. Choosing the plane $R_3=\dot{R}_3=0$ 
as the plane of motion  of the primaries and introducing polar coordinates
\begin{equation}
\begin{split}
& R_1 \equiv R \cos u,\\
& R_2 \equiv R \sin u,
\end{split}
\label{Brumberg4.16}
\end{equation}
Eq. (\ref{Brumberg4.10}) gives
\begin{equation}
\begin{split}
& \ddot{R}-R\dot{u}^2=B_1 \cos u + B_2 \sin u, \\
&\dfrac{{\rm d}}{{\rm d}t}\left(R_2 \dot{u} \right) = R \left(-B_1 \sin u + B_2 \cos u \right),
\end{split}
\label{Brumberg4.17}
\end{equation}
$B_1$ and $B_2$ having the same properties of the coefficients occurring on the right-hand side of 
(\ref{Brumberg4.10}). If we seek solutions having the form
\begin{equation}
\begin{split}
R &= A + \Delta R, \\
u &= \Lambda + \Delta u, 
\end{split}
\end{equation}
where
\begin{equation}
\Lambda = n t + \Lambda_0,
\end{equation}
($A$, $\Lambda$, $n$, and $\Lambda_0$ being constants) the correction terms $\Delta R \sim \O(c^{-5})$ 
and $\Delta u \sim \O(c^{-5})$ (which are caused by the dissipative terms occurring in 
Eq. (\ref{Brumberg4.17})) are found to be 
\begin{equation}
\begin{split}
\Delta R & = - 2 A k \Lambda, \\
\Delta u &= \dfrac{3}{2} k \Lambda^2,
\end{split}
\end{equation}
$k$ being a small parameter given by
\begin{equation}
k= \dfrac{32}{5 c^5} n^5 A^5 \mu.  
\end{equation}
Therefore, putting together all the above results, the quasi-circular orbits of the two primaries read as
\begin{equation}
\begin{split}
R &= A \left(1-2k \Lambda\right), \\
u &= \Lambda + \dfrac{3}{2} k \Lambda^2.
\label{Brumberg4.25}
\end{split}
\end{equation}
At this stage, we can consider the dynamics of a test particle immersed in the gravitational field 
of the binaries moving along the quasi-circular orbits (\ref{Brumberg4.16}) and (\ref{Brumberg4.25}). 
The equations of motion of the small planetoid are given by
\begin{equation}
\ddot{x}^i = -\dfrac{1}{2} \dfrac{\partial}{\partial x^i} h^{(2)}_{00}
= - GM_1 \dfrac{x^i-x^i_1}{(r_1)^3}-GM_2 \dfrac{x^i-x^i_2}{(r_2)^3},
\end{equation}
with
\begin{equation}
\begin{split}
r_K & \equiv \sqrt{\left(x-x_K \right)_i \left(x-x_K \right)^i}, \\
x^i_1 & \equiv \dfrac{M_2}{(M_1+M_2)}R^i, \\
x^i_2 & \equiv \dfrac{M_1}{(M_1+M_2)}R^i, 
\end{split}
\end{equation}
where the functions $R^i$ are given by Eqs. (\ref{Brumberg4.16}) and (\ref{Brumberg4.25}). Such a 
problem is called quasi-Newtonian restricted three-body problem and its main peculiarity is represented 
by the fact that radiation terms are taken into account. Like in the Newtonian case, 
the rotating synodic frame can be employed, with coordinates
\begin{equation}
\begin{split}
x^1 & \equiv \xi^1 \cos u - \xi^2 \sin u, \\
x^2 & \equiv \xi^1 \sin u + \xi^2 \cos u, \\
x^3 & \equiv \xi^3,
\end{split}
\end{equation}
where the coordinates of the binary components depend on time only through the radiation terms
\begin{equation}
\begin{split}
\xi^1_K = \pm \dfrac{M_J}{(M_1+M_2)}R, \;\;\;\; (K \neq J),
\end{split}
\end{equation} 
($K=1,2$ and $J=1,2$ labelling the primaries, the plus sign occurring if $K=1$, 
while the negative one if $K=2$) and
\begin{equation}
\xi^2_K = \xi^3_K =0.
\end{equation}
Therefore, the dynamical equations for the test body assume the form \cite{Brumberg2003}
\begin{equation}
\ddot{\xi}^1 =-\dfrac{G M_1}{(r_1)^3} \left(\xi^1-\dfrac{M_2}{(M_1+M_2)} R \right) -\dfrac{G M_2}{(r_2)^3} 
\left(\xi^1+\dfrac{M_1}{(M_1+M_2)} R \right) + 2 \dot{u} \dot{\xi}^2 +(\dot{u})^2 \xi^1+ \ddot{u} \xi^2, 
\label{Brumberg5.7}
\end{equation}
\begin{equation}
\ddot{\xi}^2= -\left[\dfrac{G M_1}{(r_1)^3}+\dfrac{G M_2}{(r_2)^3}\right]\xi^2 
-2 \dot{u} \dot{\xi}^1 +(\dot{u})^2 \xi^2- \ddot{u} \xi^1, 
\end{equation}
\begin{equation}
\ddot{\xi}^3 =  -\left[\dfrac{G M_1}{(r_1)^3}+\dfrac{G M_2}{(r_2)^3}\right]\xi^3,
\label{Brumberg5.9}
\end{equation}
with
\begin{equation}
\begin{split}
(r_1)^2 & = \left(\xi^1-\dfrac{M_2}{(M_1+M_2)}R \right)^2 + \left(\xi^2\right)^2+ \left(\xi^3\right)^2,\\
(r_2)^2 & = \left(\xi^1+\dfrac{M_1}{(M_1+M_2)}R \right)^2 + \left(\xi^2\right)^2+ \left(\xi^3\right)^2.
\end{split}
\end{equation}
We are now ready to describe the most significant features of this system. First of all, since, 
unlike the classic case, the system (\ref{Brumberg5.7})--(\ref{Brumberg5.9}) is not autonomous, the 
Jacobi integral gets replaced by the Jacobi quasi-integral relation
\begin{equation}
\begin{split}
\dfrac{1}{2} \sum_{i=1}^{3} \left(\dot{\xi}^i\right)^2 &=\dfrac{GM_1}{r_1}+\dfrac{GM_2}{r_2}
+\dfrac{1}{2} \left( \dot{u}\right)^2 \left[ \left(\xi^1\right)^2+ \left(\xi^2\right)^2 \right] \\
& + 2G(M_1+M_2)\mu n  k A \int {\rm d}t \left[\left(\xi^1-\dfrac{M_2}{(M_1+M_2)}R\right)\dfrac{1}{(r_1)^3}
-\left(\xi^1+\dfrac{M_1}{(M_1+M_2)}R\right)\dfrac{1}{(r_2)^3}\right] \\
&+ \ddot{u} \int {\rm d}t \left[ \xi^2 \left(\dot{\xi}^1-\dot{u}\xi^2 \right) -\xi^1 
\left(\dot{\xi}^2+\dot{u}\xi^1 \right)\right]-\dfrac{1}{2}C_{J},
\end{split}
\end{equation}
$C_J$ being the classical Jacobi constant. Moreover, we can evaluate the relativistic phenomena affecting 
the classical position of Lagrangian points. In fact, the Newtonian theory predicts for the circular 
restricted three-body problem five equilibrium points which can be divided into two categories: 
the unstable collinear ones ($L_1$, $L_2$, and $L_3$) lying on the line joining the primaries and 
the triangular ones ($L_4$ and $L_5$) being the vertices of an equilateral triangle whose basis is 
represented by the line connecting the massive bodies. Thus, by analogy in the quasi-Newtonian 
problem stated above there exist five quasi-libration points.

We focus on the non-collinear points. Classically, for $k=0$ the position of $L_4$ is given 
by (in the planar case where $\xi^3_{\;cl}=0$)
\begin{equation}
\begin{split}
\xi^1_{\;cl} & = \dfrac{M_2-M_1}{2(M_1+M_2)}A, \\
\xi^2_{\;cl} &= \dfrac{\sqrt{3}}{2}A ,\\
r_{1,cl}&=r_{2,cl}=A.
\end{split}
\end{equation}
By employing the post-Newtonian pattern, we can look for coordinates of the 
quasi-Lagrangian point $L_4$ having the form
\begin{equation}
\begin{split}
\xi^1=\xi^1_{\;cl} + \delta\xi^1,\\
\xi^2= \xi^2_{\;cl} + \delta \xi^2 ,\\
\xi^3= \delta \xi^3,
\end{split}
\end{equation}
and
\begin{equation}
\begin{split}
r_1 &= R \left(1-\dfrac{1}{2R} \delta \xi^1 +\dfrac{\sqrt{3}}{2} \delta \xi^2 + \dots \right),\\
r_2 &= R \left(1+\dfrac{1}{2R} \delta \xi^1 +\dfrac{\sqrt{3}}{2} \delta \xi^2 + \dots \right).
\end{split}
\end{equation}
By substituting the above ansatz into Eqs. (\ref{Brumberg5.7})--(\ref{Brumberg5.9}) 
we obtain, to first order in $k$, 
\begin{equation}
\begin{split}
\delta \xi^1&=\dfrac{2}{3\sqrt{3}}\dfrac{1-\mu}{\mu} Ak, \\
\delta \xi^2 &= -\dfrac{2}{9}\dfrac{M_2-M_1}{\mu(M_1+M_2)}Ak,\\
\delta \xi^3 &=0,
\end{split}
\end{equation}
showing that the {\it binary} gravitational radiation terms force the coordinate of $L_4$ to change 
with time, hence revealing its true quasi-equilibrium nature\footnote{These are {\it indirect} 
perturbations affecting the position of libration points caused by radiation terms occuring in 
Eq. (\ref{Brumberg4.25}).}. The same arguments can be applied also to the classical collinear 
Lagrangian points. Again, their coordinates are no longer constant in time and they are turned into 
quasi-libration points whose positions depend secularly on time \cite{Brumberg2003}.

\section{Potentialities of the planetary ephemeris program PEP}

An approach to the test of the foundations of general relativistic celestial mechanics different 
from, and complementary to, what reported in this work is the analysis of solar-system metric data
carried out with the Planetary Ephemeris Program, hereafter referred to as PEP. This is a software
package developed at the Harvard-Smithsonian Center for Astrophysics over the past several decades
\cite{Reasenberg1975,Reasenberg1980}. PEP has been used successfully to describe the three-body
problem of the Sun-Earth-Moon in the weak-field slow-motion regime, in the solar system
barycenter frame, taking into account numerically additional perturbing effects (the other planets,
Pluto, asteroids, Earth/Moon geodesy effects). In particular, the de Sitter precession of the
orbit of the Moon has been measured \cite{Reasenberg1987,Shapiro1988,Chandler1988} and this,
as well as other general relativity tests (equivalence principle etc.) are being continuously
refined \cite{Williams2004,Williams2009, MartiniDellAgnello2016} by using lunar laser ranging
metric data \cite{Murphy2013}. Even this experimental approach is affected by technical
complexities of the underlying models and of the code for implementing the models 
\cite{Reasenberg2016}. Integrated analytical-numerical approaches (this work and PEP's), strongly
aided by an ever-increasing set of experimental metric measurements from new solar system 
missions (on the Moon, but also on Mars \cite{DellAgnello2017}), have the potential to improve
our detailed knowledge of the solar system dynamics and of how to use it to probe general relativity. 

\section{Concluding remarks and open problems}

In the very accurate review paper in Ref. \cite{Brumberg2013}, the author stresses that 
relativistic celestial mechanics has one irrefutable merit, i.e., its exceptionally high
precision of observations absolutely unattainable in cosmology and astrophysics. In his opinion,
with which we agree, the final goal of relativistic celestial mechanics is to answer the
question whether general relativity alone is able of accounting for all observed motions of
celestial bodies and the propagation of light. The work in Ref. \cite{Brumberg2013} lists
eventually the following major tasks of relativistic celestial mechanics in the years to come:
the investigation of general relativistic equations of motion, orbital evolution with
emission of gravitational radiation, general relativistic treatment of body rotation, and the
motion of bodies in the background of the expanding universe. 
 
In particular, the most important unsolved problem of Newtonian 
celestial mechanics, which is possibly the proof
of stability of the solar system (cf. the work in Refs. \cite{LAS,BAT,Pinzari}), remains an
unsettled issue also in general relativity, but, well before achieving this goal, it is important
to improve the tools at our disposal in order to perform at least numerical simulations of planetary
and satellite motion when all celestial bodies (including moons, asteroids and comets) are taken
into account. Mankind has now at disposal some high performance supercomputers with capabilities
that were inconceivable in the nineties, but unless we succeed in overcoming technical difficulties
resulting from non-local terms in the equations of motion, we will have to limit ourselves to the
spectacular progress achieved in the investigation of two-body dynamics in strong fields
\cite{DA1,DA2,DA3,DA4,ALBA2,DA5,DA6,DA7,DA8,DA9,DA10,DA11,DA12,DA13,DA14,DA15}. Hopefully,
the weak-gravity regime that is relevant for many aspects of solar system dynamics will be
better understood as well in the (near) future. 
 
\acknowledgments
The authors are grateful to the Dipartimento di Fisica ``Ettore Pancini'' of Federico II University for
hospitality and support. Their work has been supported by the 
INFN funding of the NEWREFLECTIONS experiment.

\end{document}